\documentclass[11pt]{article}
\usepackage{amsmath,amssymb,color,epsf}
\usepackage{amssymb,slashed,latexsym,amsmath,multirow,color}
\usepackage{graphics}
\makeatletter
\@addtoreset{equation}{section}
\makeatother

\usepackage{amsfonts}

\newcommand{\hoch}[1]{$\, ^{#1}$}

\newcommand{\be}{\begin{equation}}
\newcommand{\ee}{\end{equation}}
\newcommand{\bea}{\setlength\arraycolsep{2pt} \begin{eqnarray}}
\newcommand{\eea}{\end{eqnarray}}
\newcommand{\nn}{\nonumber}

\usepackage[font=footnotesize,labelsep=newline,labelfont=sc,justification=centering,position=top]{caption}

\newcommand{\D}{\mathcal{D}}

\usepackage{hyperref}

\newcommand{\ft}[2]{{\textstyle\frac{#1}{#2}}}
\def\rmi{{\rm i}}
\newsavebox{\uuunit}
\sbox{\uuunit}
    {\setlength{\unitlength}{0.825em}
     \begin{picture}(0.6,0.7)
        \thinlines
        \put(0,0){\line(1,0){0.5}}
        \put(0.15,0){\line(0,1){0.7}}
        \put(0.35,0){\line(0,1){0.8}}
       \multiput(0.3,0.8)(-0.04,-0.02){12}{\rule{0.5pt}{0.5pt}}
     \end {picture}}

\newcommand{\SU}{\mathop{\rm SU}}

\def\be{\begin{equation}}
\def\ee{\end{equation}}
\def\ba{\begin{array}}
\def\ea{\end{array}}
\def\bea{\begin{eqnarray}}
\def\eea{\end{eqnarray}}
\def\bd{\begin{displaymath}}
\def\ed{\end{displaymath}}

\def\nn{\nonumber}

\def\a{\alpha}
\def\b{\beta}
\def\g{\gamma}

\def\d{\delta}
\def\D{\Delta}
\def\e{\epsilon}
\def\ve{\varepsilon}
\def\f{\phi}
\def\vf{\varphi}

\def\p{\psi}

\def\l{\lambda}
\def\L{\Lambda}
\def\m{\mu}
\def\n{\nu}
\def\r{\rho}
\def\s{\sigma}

\def\t{\tau}

\def\o{\omega}

\def\nn{\nonumber}
\def\cD{\mathcal{D}}
\def\cN{\mathcal{N}}

\def\cL{\mathcal{L}}

\begin{document}

\begin{flushright}
\hfill{ \
\ \ \ \ MIFPA-13-01\ \ \ \ }
\end{flushright}
\vskip 1.2cm
\begin{center}
{\Large \bf Supersymmetric Completion of Gauss-Bonnet Combination
 }
\\
{\Large \bf in Five Dimensions}

\end{center}
\vspace{25pt}
\begin{center}
{\Large {\bf }}

\vspace{10pt}

{\Large Mehmet Ozkan\hoch{} and Yi Pang\hoch{}
}

\vspace{10pt}

\hoch{} {\it George P. \& Cynthia Woods Mitchell  Institute
for Fundamental Physics and Astronomy,\\
Texas A\&M University, College Station, TX 77843, USA}

\vspace{10pt}

\underline{ABSTRACT}

\end{center}

Based on superconformal tensor calculus in five dimensions, we construct the
supersymmetric completion of Gauss-Bonnet combination. We study the vacuum solutions with
$AdS_2\times S^3$ and $AdS_3\times S^2$ structures. We also analyze the spectrum around a maximally
supersymmetric ${\rm Minkowski}_5$.

\vspace{15pt}

 \vfill

\hrule width 3.cm \vspace{2mm}{\small \noindent e-mails:\, mozkan@tamu.edu,\, pangyi1@physics.tamu.edu}
\thispagestyle{empty}

\thispagestyle{empty}
\voffset=-40pt

\newpage

\tableofcontents


\newpage


\section{Introduction}

Higher-curvature corrections to the Einstein-Hilbert action naturally arise in the low-energy limit of string theories and play an important role in their compactification \cite{Candelas:1985en,Zwiebach:1985uq} where curvature squared terms can appear in the lower dimensional effective action. In this context, higher-curvature corrections take the form of an infinite series required by on-shell supersymmetry which only works order by order. Many attempts have been carried out in the explicit construction of supersymmetric higher-derivative terms. For instance, supersymmetric $R^2$ terms were studied in \cite{Romans:1985xd}-\cite{Fre:1991ef} motivated by supersymmetrizing the Lorentz Chern-Simons term that is indispensable to the anomaly cancelation \cite{Green:1984sg}.  However, if the higher-curvature terms are treated as perturbative interactions leaving the degrees of freedom and propagator unchanged, only the coefficient in front of $R^{\mu\nu\rho\sigma}R_{\mu\nu\rho\sigma}$ has a definite meaning, since a field redefinition of the form
\be
g_{\mu\nu}' = g_{\mu\nu} + a R_{\mu\nu} + b g_{\mu\nu} R
\ee
can shift the coefficients in front of the $R^{\mu\nu}R_{\mu\nu}$ and $R^2$ terms to arbitrary values \cite{Deser:1986xr}. On the other hand,
there are also situations where it is interesting to consider a finite number of higher-curvature terms on the same footing as Einstein-Hilbert term, since
higher-derivative terms can improve the ultraviolet behavior of gravitational theories \cite{Stelle:1976gc}. Among all the quadratic curvature theories of gravity, the Gauss-Bonnet combination is singled out since it is ghost-free, sharing the similar property with Einstein gravity. Its form is given by
\be
e^{-1} \cL_{{\rm GB}} = R_{\m\n\r\s} R^{\m\n\r\s} - 4 R_{\m\n} R^{\m\n} + R^2.
\label{GBCombination}
\ee
In dimensions $D\leq6$, certain types of off-shell formulation of supergravity are known in which the higher-derivative bosonic terms can be extended to complete and independent super-invariants with only a finite number of terms being required. Progresses on supersymmetrizing the Gauss-Bonnet combination have been made. In four-dimensional $\cN =1$ supergravity, supersymmetric Gauss-Bonnet term with matter coupling was constructed in \cite{Cecotti:1987mr,Ferrara:1988pd,BuchKuz,Ferrara:1988qx,LeDu:1997us}; in six-dimensional chiral $\cN =2$ supergravity, partial results on the Gauss-Bonnet super-invariant were given in \cite{Bergshoeff:1986vy,Bergshoeff:1986wc}.

In this work, we study the supersymmetric completion of Gauss-Bonnet combination in five dimensions. We use the five-dimensional superconformal tensor calculus \cite{Kugo:2000hn,Fujita:2001kv} which facilitates the construction tremendously. Since superconformal tensor calculus is an off-shell formalism, the analysis of the higher derivative terms can be done without modifying the supersymmetry transformation rules. The off-shell nature of the supersymmetric invariants allow us to combine different invariants to obtain more general theories.

The crucial observation in our construction of supersymmetric Gauss-Bonnet combination is that although three independent curvature squared terms enter the expression of Gauss-Bonnet combination, such an off-shell construction might be possible with only two independent curvature squared super-invariants. This observation is based on the fact that the Riemann squared invariant obtained in \cite{Bergshoeff:2011xn} using the Dilaton Weyl multiplet contains
an ordinary kinetic term for the auxiliary vector field $V_{\mu}^{ij}$. Thus the Riemann square extended
Poincar\'e supergravity contains a dynamical massive auxiliary vector in its spectrum which forms the same multiplet with the massive graviton generated by the Riemann squared term. By counting degrees of freedom, we notice that it might always be the case (except for the pure Ricci scalar squared invariant) that when formulated in terms of Dilaton Weyl multiplet, the curvature squared super-invariant includes an ordinary kinetic term for the auxiliary vector field $V_{\mu}^{ij}$. Therefore, if there exist two independent curvature squared super-invariants, a particular combination of them can be formed in which the kinetic term for the auxiliary vector vanishes. This implies that there is no massive graviton since the massive vector and massive graviton fall into the same multiplet, suggesting that the curvature squared terms comprise Gauss-Bonnet combination.

Based on the above observation, we start looking for another curvature squared invariant constructed in terms of the Dilaton Weyl multiplet besides the known Riemann tensor squared  invariant. An obvious candidate is $\widehat{C}^{\mu\nu\rho\sigma}\widehat{C}_{\mu\nu\rho\sigma}$, which is the superconformal extension of the Weyl tensor squared term whose supersymmetric completion was obtained previously in \cite{Hanaki:2006pj}\footnote{The on-shell theory of this model is derived in \cite{Cremonini:2008tw}.} using the Standard Weyl multiplet coupled to the vector multiplet. Utilizing superconformal tensor calculus, we supersymmetrize the square of super-covariant Weyl tensor. We find that in addition to the Weyl tensor squared term, the bosonic action acquires a Ricci scalar squared term arising from the square of $D$, which is a fundamental scalar field in the Standard Weyl multiplet but a composite field in the Dilaton Weyl multiplet. Equivalently, the curvature squared terms in the action take the form of $C^{\mu\nu\rho\sigma} C_{\mu\nu\rho\sigma} + \ft16 R^2$. Since
\be
C^{\mu\nu\rho\sigma} C_{\mu\nu\rho\sigma} + \ft16 R^2=R^{\mu\nu\rho\sigma} R_{\mu\nu\rho\sigma}-\ft43R^{\mu\nu}R_{\mu\nu}+\ft13R^2
\ee
the ratio of coefficients in front of the $R^{\mu\nu}R_{\mu\nu}$ and $R^2$ terms is -4, which is the required value to obtain the supersymmetric completion of Gauss-Bonnet invariant by combining a Riemann squared invariant with an appropriate coefficient.

This paper is organized as follows. In section \ref{s: SCMultplets}, we briefly review the superconformal multiplets of five-dimensional supergravity constructed in \cite{Fujita:2001kv, Bergshoeff:2001hc, Bergshoeff:2002qk}. In section \ref{s: SCActions}, we review the construction of the superconformal linear multiplet action \cite{Coomans:2012cf} and obtain a superconformal action for the Yang-Mills multiplet coupled to the Dilaton Weyl multiplet. In section \ref{s: GaugeFix}, we fix the superconformal symmetries to obtain an off-shell Poincar\'e supergravity and an off-shell Yang-Mills theory coupled to the Dilaton Weyl multiplet. Using a map between the Yang-Mills multiplet and the Dilaton Weyl multiplet \cite{Bergshoeff:2011xn}, we reconstruct the off-shell supersymmetric Riemann squared action. In section \ref{s: SCGB}, we present the supersymmetric completion of $C^{\mu\nu\rho\sigma} C_{\mu\nu\rho\sigma} + \ft16 R^2$ and combine it with a supersymmetric Riemann squared invariant to obtain the supersymmetric Gauss-Bonnet combination. In section 6, we derive the $D=5$ on-shell minimal Einstein Hilbert supergravity and on-shell Gauss-Bonnet extended Einstein-Maxwell supergravity. In section \ref{s: Vacuum}, we discuss the vacuum solutions with $AdS_3\times S^2$ and $AdS_2\times S^3$ structures. The bosonic spectrum around a maximally supersymmetric Minkowski${}_5$ vacuum is also analyzed. In section \ref{s:Conc} we give conclusion and discussions.


\section{Superconformal Multiplets}\label{s: SCMultplets}

In this section, we introduce the basic elements of five dimensional superconformal tensor calculus with eight supercharges \cite{Fujita:2001kv, Bergshoeff:2001hc}.  In section 2.1, we present the Dilaton Weyl multiplet adopted in our construction. In the subsequent two subsections, we briefly review two superconformal  matter multiplets of  $D=5, \, \cN = 2$ theory: the Yang-Mills multiplet and the linear multiplet, which are used as compensator multiplets in the construction of superconformal actions.

\subsection{Dilaton Weyl Multiplet}\label{ss: DWM}

In \cite{Bergshoeff:2001hc}, it was established that there exist two different Weyl multiplets for $\cN =2$  conformal supergravity in five dimensions: the Standard Weyl multiplet and the Dilaton Weyl multiplet. These two multiplets have the same contents of gauge fields but different matter fields. However, the matter fields of the Standard multiplet can be built from the fundamental fields in the Dilation Weyl multiplet as composite fields. The gauge sector of the Dilaton Weyl multiplet consists of a f\"unfbein $e_\m{}^a$, a gravitino, $\p_\m{}^i$, the dilatation gauge field $b_\m$, and the $\SU(2)$ gauge field $V_\m^{ij}$. Since these gauge fields account for $21$(bosonic) $+ 24$ (fermionic) degrees of freedom, they cannot form a super-multiplet. Therefore, matter fields are needed to comprise a superconformal Weyl multiplet. For the Dilaton Weyl multiplet, the matter sector consists of a physical vector $C_\m$, an antisymmetric two-form gauge field $B_{\m\n}$, a dilaton field $\s$ and a dilatino $\p^i$. The $Q$, $S$ and $K$ transformation rules for the Dilaton Weyl Multipet are given by \cite{Bergshoeff:2001hc}
\bea
\d e_\m{}^a   &=&  \ft 12\bar\e \g^a \psi_\m  \nn\, ,\\
\d \psi_\m^i   &=& (\partial_\m+\tfrac{1}{2}b_\m+\tfrac{1}{4}\omega_\m{}^{ab}\g_{ab})\e^i-V_\m^{ij}\e_j + \rmi \g\cdot {T} \g_\m
\e^i - \rmi \g_\m
\eta^i  \nn\, ,\\
\d V_\m{}^{ij} &=&  -\ft32\rmi \bar\e^{(i} \phi_\m^{j)} +4
\bar\e^{(i}\g_\m {\chi}^{j)}
  + \rmi \bar\e^{(i} \g\cdot {T} \psi_\m^{j)} + \ft32\rmi
\bar\eta^{(i}\psi_\m^{j)} \nn\, ,\\
 \d C_\m
&=& -\ft12\rmi \s \bar{\e} \p_\m + \ft12
\bar{\e} \g_\m \p, \nonumber\\
 \d B_{\m\n}
&=& \ft12 \s^2 \bar{\e} \g_{[\m} \p_{\n]} + \ft12 \rmi \s \bar{\e}
\g_{\m\n} \p + C_{[\m} \d(\e) C_{\n]}, \nonumber\\
\d \p^i &=& - \ft14 \g \cdot \widehat{G} \e^i -\ft12\rmi \slashed{\mathcal{D}} \s
\e^i + \s \g \cdot {T} \e^i -\ft14\rmi\s^{-1}\e_j \bar\p^i \p^j  + \s\eta^i \,,\nonumber\\
\d \s &=& \ft12 \rmi \bar{\e} \p \, ,\nonumber\\
 \d b_\m       &=& \ft12 \rmi \bar\e\phi_\m -2 \bar\e\g_\mu {\chi} +
\ft12\rmi \bar\eta\psi_\mu+2\Lambda _{K\mu } \,,
\label{TransDW}
\eea
where
\begin{eqnarray}
\mathcal{D}_\mu\, \s &=&  (\partial_\mu - b_\mu) \s
- \tfrac12\, \rmi\bar{\psi}_\mu \p \ ,
\nn\\
\mathcal{D}_\mu \p^i &=&  (\partial_\mu -\ft32 b_\mu +\ft14\,  \o_\mu{}^{ab}\g_{ab} ) \p^{i} - V_\mu^{ij} \p_j +\tfrac 14 \g
\cdot \widehat{G} \p_\mu^i  \nn\\
&& + \ft12\rmi \slashed{D} \s \p_\mu^i
 +\ft14\rmi\s^{-1}\p_{\m j}\bar\p^i\p^j  - \s  \g \cdot T \p_\mu^i - \s
\phi_\mu^i\,,
\label{cd2}
\end{eqnarray}
and the supercovariant curvatures are defined according to
\bea
\widehat{G}_{\m\n}  &=& G_{\m\n} - \bar{\p}_{[\m} \g_{\n]} \p + \tfrac 12 \rmi
\s \bar{\p}_{[\m} \p_{\n]} \label{hatG} ,\nn\\
\widehat{H}_{\m\n\r} &=& H_{\m\n\r} - \ft34
\s^2 \bar{\p}_{[\m} \g_\n \p_{\r]} - \ft32\rmi \s \bar{\p}_{[\m}
\g_{\n\r]} \p,
\label{DefH}
\eea
In above expressions, $G_{\m\n}=2 \partial_{[\mu } C_{\nu ]}$ and $H_{\m\n\r} = 3\partial _{[\mu }B_{\nu \rho ]} + \ft32 C_{[\m}  G_{\n\r]}$. Note that $\widehat{G}_{\m\n}$ and ${\widehat H}_{\mu\nu\rho}$ are invariant under following gauge transformations
\be \delta C_\mu= \partial_\mu\Lambda\ ,\qquad \delta B_{\mu\nu} =
2\partial_{[\mu} \Lambda_{\nu]} -\ft12 \Lambda G_{\mu\nu}.
\ee
The definitions of spin connection $\o_\m{}^{ab}$ and the $S$-supersymmetry gauge field $\phi_\m^i$ are given in \cite{Bergshoeff:2001hc}
\bea
\o_\m{}^{ab} &=& 2 e^{\n[a}\partial_{[\m} e_{\n]}{}^{b]} - e^{\n[a} e^{b]\s} e_{\m c} \partial_\n e_\s{}^c + 2 e_\m{}^{[a} b^{b]} + \ft12 \bar\p^{[a} \g^{b]}\p_\m + \ft14 \bar\p^a \g_\m \p^b ,\nn\\
\phi_\m^i &=& \ft13 \rmi \g^a \widehat{R}^{'}_{\m a}{}^{i}(Q) - \ft1{24} \rmi \g_\m \g^{ab} \widehat{R}^{'}_{ab}{}^{i}(Q),
\eea
where $\widehat{R}_{\m\n}^i(Q)$, the supercovariant curvature of gravitino is defines as \cite{Bergshoeff:2001hc}
\bea
\widehat{R}_{\m\n}^i(Q) &=&\widehat{R}{'}_{\m\n}{}^i(Q) -2i\g_{[\m}\phi_{\n]}^i \,,\nn\\
\widehat{R}{'}_{\m\n}{}^i(Q) &=& 2\partial_{[\m}\p_{\n]}^i+\frac{1}{2}\o_{[\m}{}^{ab}\g_{ab}\p_{\n]}^i+b_{[\m}\p_{\n]}^i-2V_{[\m}^{ij}\p_{\n] j} +2i\g\cdot T\g_{[\m}\p_{\n]}^i.
\eea
For future reference, we also give the supercovariant curvature of $\o_\m{}^{ab}$ and $V^{ij}_{\mu}$ \cite{Bergshoeff:2001hc}
\bea
\widehat{R}_{\m\n}{}^{ab}(M)&=&2\partial_{[\m}\o_{\n ]}{}^{ab}+2\o_{[\m}{}^{ac}\o_{\n ]c}{}^{b} + 8 f_{[\m}{}^{[a}e_{\n ]}{}^{b]}+\rmi \bar\p_{[\m}\g^{ab}\p_{\n ]} + \rmi \bar\p_{[\m}\g^{[a} \g \cdot T \g^{b]}\p_{\n ]}  \nn\\
&& +\bar\p_{[\m} \g^{[a} \widehat{R}_{\n ]}{}^{b]}(Q)+\tfrac12 \bar\p_{[\m}\g_{\n ]} \widehat{R}^{ab}(Q) -8 \bar\p_{[\m} e_{\n ]}{}^{[a} \g^{b]}\chi+i\bar{\phi}_{[\m} \g^{ab} \p_{\n]},\cr
\widehat{R}_{\m\n}{}^{ij}(V)&=&2\partial_{[\m} V_{\n]}{}^{ij} -2V_{[\m}{}^{k( i}
V_{\n ]\,k}{}^{j)} {-3\rmi}{\bar\f}^{( i}_{[\m}\p^{j)}_{\n ]}  - 8 \bar{\p}^{(i}_{[\m}
\g_{\n]} \chi^{j)} - \rmi \bar{\p}^{(i}_{[\m} \g\cdot T \psi_{\n]}^{j)}.
\label{RM}
\eea
The $Q$- and $S$- transformations of the field strengths $\widehat{G}_{\m\n}$ and $\widehat{H}_{abc}$ are presented in \cite{Bergshoeff:2001hc}
\bea
\d \widehat{G}_{\m\n} &=& - \tfrac12 \rmi \s \bar\e \widehat{R}_{\m\n}(Q) - \bar\e \g_{[\m} {\cal{D}}_{\n ]}\p + \rmi \bar\e \g_{[\m} \g \cdot T \g_{\n ]} \p + \rmi \bar\eta \g_{\m\n} \p \,, \nn\\
\d \widehat{H}_{abc} &=& -\ft34 \s^2 \bar\e \g_{[a} \widehat{R}_{bc]}(Q) + \ft32 \rmi \bar\e \g_{[ab} {\mathcal{D}}_{c]} \p + \ft32 \rmi \bar\e \g_{[ab} \p {\mathcal{D}}_{c]} \s\nn\\
&& -\ft32 \s \bar\e \g_{[a} \g \cdot T \g_{bc]} \p -\ft32 \bar\e \g_{[a} \widehat{G}_{bc]} \p - \ft32 \s \bar\eta \g_{abc} \p\,.
\eea
The expressions for the composite fields $T_{ab}, \chi^i$ and $D$ are given as follows \cite{Bergshoeff:2001hc}
\bea
T_{ab} &=& \ft18 \s^{-2} \Big( \s \widehat G_{ab} + \ft16 \e_{abcde} \widehat H^{cde} + \ft14 \rmi \bar\p \g_{ab} \p \Big) \,, \nn\\
\chi^i &=& \ft18 \rmi \s^{-1} \slashed{\cD} \p^i + \ft1{16} \rmi \s^{-2} \slashed{\cD} \s \p^i - \ft1{32} \s^{-2} \g \cdot \widehat G \p^i + \ft14 \s^{-1} \g \cdot {T} \p^i \nn\\
&& + \ft1{32} \rmi \s^{-3} \p_j \bar\p^i\p^j ,\,\nn\\
D &=& \ft14 \s^{-1} \Box^c \s + \ft18 \s^{-2} (\cD_a \s) (\cD^a \s) - \ft1{16} \s^{-2} \widehat G_{\m\n} \widehat G^{\m\n}\nn\\
&& - \ft18 \s^{-2} \bar\p \slashed{\cD} \p  -\ft1{64} \s^{-4} \bar\p^i \p^j \bar\p_i \p_j - 4 \rmi \s^{-1} \p {\chi} \nn\\
&& + \Big( - \ft{26}3 {T_{ab}} + 2 \s^{-1} \widehat G_{ab} + \ft14 \rmi \s^{-2} \bar\p \g_{ab} \p \Big) {T}^{ab}\,,
\label{UMap}
\eea
where the superconformal d'Alambertian for $\s$ is given by
\bea
&&\Box^c \s= (\partial^a - 2b^a + \o_b{}^{ba}) \cD_a \s - \ft12 \rmi \bar\p_a \cD^a\psi - 2\sigma \bar\p_a \g^a {\chi} \nn\\
&&\quad\quad + \ft12 \bar\p_a \g^a \g \cdot {T} \psi + \ft12 \bar\phi_a \g^a \psi + 2 f_a{}^a \sigma,\cr
&&f^a_\m=  - \ft16{\cal R}_\mu {}^a + \ft1{48}e_\mu {}^a {\cal R},\quad{\cal R}_{\mu \nu }\equiv \widehat{R}_{\mu \rho }^{\prime~~ab}(M) e_b{}^\rho
e_{\nu a},\quad{\cal R}\equiv {\cal R}_\mu {}^\mu.
\eea
The notation $\widehat{R}'(M)$ indicates that we have omitted the $f_\m{}^a$ term in $\widehat{R}(M)$.
It was established in \cite{Bergshoeff:2001hc} that one can also construct another Weyl multiplet, the Standard Weyl multiplet if considering $T_{ab} , D$ and $\chi^i$ as fundamental fields instead as a matter sector in addition to the gauge sector of the Weyl multiplet. Therefore, the composite expressions of these fields establish a map from the Dilaton Weyl multiplet to the Standard Weyl multiplet. We refer to \cite{Bergshoeff:2001hc, Coomans:2012cf} for readers interested in the derivation of this map and the five-dimensional Weyl multiplets in superconformal theory. For later convenience, we also present the $Q$- and $S$- transformations of the composite fields \cite{Bergshoeff:2001hc}
\bea
\d T_{ab} &=& \tfrac12 \rmi\bar\e \g_{ab} \chi - \tfrac3{32} \rmi \bar\e \widehat{R}_{ab}(Q)\,, \cr
\d \chi^i &=& \tfrac14 \e^i D - \tfrac1{64} \g \cdot  \widehat{R}^{ij}(V) \e_j + \tfrac18 \rmi \g^{ab}\slashed{\mathcal{D}}T_{ab}\e^i - \tfrac18 \rmi \g^a  \mathcal{D}^b T_{ab} \e^i \nn\\
&& - \tfrac14 \g^{abcd}T_{ab} T_{cd} \e^i + \tfrac16 T^2 \e^i + \tfrac14 \g \cdot T \eta^i\,, \nn\\
\d D &=& \bar\e \slashed{\mathcal{D}}\chi - \tfrac53 \rmi \bar\e \g \cdot T \chi - \rmi \bar\eta \chi\,,
\label{ComposiT}
\eea
where the supercovariant derivatives of the composite fields are
\bea
\mathcal{D}_\m\chi^i&=&(\partial_\mu - \tfrac72 b_\mu +\tfrac14 \omega_\mu{}^{ab} \g_{ab})\chi^i -V_\mu^{ij}\chi_j
 - \tfrac14 \p_\m^i D  + \tfrac1{64} \g \cdot  \widehat{R}^{ij}(V) \p_{\m j}\nn\\
&& - \tfrac18 \rmi \g^{ab}\slashed{\mathcal{D}}T_{ab}\p_\m^i + \tfrac18 \rmi \g^a  \mathcal{D}^b T_{ab} \p_\m^i  + \tfrac14 \g^{abcd}T_{ab} T_{cd} \p_\m^i - \tfrac16 T^2 \p_\m^i - \tfrac14 \g \cdot T \phi_\m^i\,, \nn \\
\mathcal{D}_\m T_{ab}&=&\partial_\m T_{ab}-b_\m T_{ab}-2\omega_\mu{}^c{}_{[a}T_{b]c}-\tfrac{1}{2}i\bar{\p}_\m\g_{ab}\chi+\tfrac{3}{32}i\bar{\p}_\m\widehat{R}_{ab}(Q)\,.
\eea

\subsection{Yang-Mills Multiplet}

The off-shell non-abelian $D=5$, $\mathcal{N}=2$ vector multiplet
consists of $8 n$ (bosonic) + $8 n$ ( fermionic) degrees of freedom
(where $n$ is the dimension of the gauge group). Denoting the
Yang-Mills index by $I$ ($I=1,\cdots,n$), the bosonic sector
consists of vector fields $A_\m^I$, scalar fields $\rho^I$ and
$\SU(2)$-triplet auxiliary fields $Y^{ij\, I} = Y^{(ij)\, I}$. $\SU(2)$-doublet fields
$\l^{i I}$ constitute the fermionic sector.

In the
background of the Dilaton Weyl multiplet, the $Q$- and $S$-transformations of the fields in the vector multiplet are given by \cite{Bergshoeff:2002qk}
\begin{eqnarray}
\d A_\m^I &=& -\ft12\rmi \rho^I \bar{\e} \p_\m + \ft12 \bar{\e}
\g_\m \lambda^I \ ,
\nonumber \\
\d Y^{ij \, I} &=& -\ft12
\bar{\e}^{(i} {\slashed{\cD}} \lambda^{j) I} + \ft12 \rmi \bar{\e}^{(i}
\g \cdot T \lambda^{j) I} - 4 \rmi \rho^I \bar{\e}^{(i} \chi^{j)} +
\ft12 \rmi \bar{\eta}^{(i} \lambda^{j) I} - \ft12 \rmi g
\bar{\e}^{(i} f_{JK}{}^I \rho^J \lambda^{j)K} \ ,
\nonumber \\
\d\lambda^{i I} &=& - \ft14 \g \cdot \widehat{F}^I \e^i -\ft12\rmi
{\slashed{\cD}}\rho^I \e^i + \rho^I \g \cdot T \e^i - Y^{ij\, I} \e_j +
\rho^I \eta^i \ ,
\nonumber \\
\d \rho^I &=& \ft12 \rmi \bar{\e}
\lambda^I \ . \label{vectconform}
\end{eqnarray}
The superconformally covariant derivatives used here are
\begin{eqnarray}
{\cD}_\mu\, \rho^I &=&
(\partial_\mu - b_\mu) \rho^I+ g f_{JK}{}^I A_\mu^J \rho^K
- \frac12\, \rmi\bar{\psi}_\mu \lambda^I \ ,
\label{deriv1}\\
{\cD}_\mu \lambda^{iI} &=&
(\partial_\mu -\ft32 b_\mu +\ft14\,  {\o}_\mu
{}^{ab}\g_{ab} ) \l^{iI} - V_\mu^{ij} \l_j^I + g f_{JK}{}^I A_\mu^J
\lambda^{iK}
\nn\\
&& +\frac 14 \g
\cdot \widehat{F}^I \p_\mu^i + \ft12\rmi \widehat{\slashed{D}}\rho^I \p_\mu^i
 + Y^{ijI} \p_{\mu\, j}- \rho^I \g \cdot T \p_\mu^i - \rho^I
\phi_\mu^i \label{deriv2} \ ,
\label{YangMills}
\end{eqnarray}
where the supercovariant Yang-Mills curvature is given as
\begin{equation}
\widehat{F}_{\m\n}^I = 2 \partial_{[\mu } A_{\nu ]}^I + g f_{JK}{}^I
A_\mu^J A_\nu^K - \bar{\p}_{[\m} \g_{\n]} \lambda^I + \frac 12 \rmi
\rho^I \bar{\p}_{[\m} \p_{\n]}\ .
\label{hatF}
\end{equation}


\subsection{Linear Multiplet}

The off-shell $D=5, {\cal{N}} = 2$ linear multiplet contains $8$ (bosonic)+$8$ (fermionic) degrees of freedom carried by the following fields
\be
(L^{ij},E^a,N,\varphi^i).
\ee
The bosonic fields are an $\SU(2)$ triplet $L^{ij} = L^{(ij)}$, a constrained vector $E_{a}$ and a scalar $N$. The fermionic fields are given by an $\SU(2)$ doublet
$\varphi^i$. In the background of the Standard Weyl multiplet, the $Q$- and $S$- transformations of the fields in the linear multiplet  are given by \cite{Coomans:2012cf}
\begin{eqnarray}
\delta L^{ij} &=& i\bar{\e}^{(i}\varphi^{j)}\,,\nn\\
\delta \varphi^{i} &=& - \tfrac{1}{2} i \slashed{\mathcal{D}}L^{ij}\e_{j} - \tfrac{1}{2} i \gamma^{a} E_{a} \e^i + \tfrac{1}{2} N \e^{i}  - \g \cdot T L^{ij} \e_{j} + 3 L^{ij}\eta_{j}\,,\nn\\
\delta E_{a} &=& -\tfrac{1}{2} i \bar{\e} \g_{ab} \mathcal{D}^{b} \varphi  - 2 \bar{\e} \gamma^{b} \varphi T_{ba} - 2\bar{\eta} \g_{a} \varphi\,, \nn\\
\delta N &=& \tfrac{1}{2} \bar{\e} \slashed{\mathcal{D}}\varphi  + \tfrac{3}{2} i \bar{\e} \g \cdot T \vf + 4 i \bar{\e}^i \chi^j L_{ij} + \tfrac{3}{2} i \bar{\eta} \varphi\,,
\label{trlm}
\end{eqnarray}
where the super-covariant derivatives are defined as
\bea
\mathcal{D}_\m L^{ij}&=&(\partial_{\m}-3b_{\m})L^{ij}+2V_{\m}{}^{(i}{}_kL^{j)k}-i\bar{\p}_{\m}^{(i}\varphi^{j)}\,, \nn \\
\mathcal{D}_\m \varphi^i&=&(\partial_{\m}-\tfrac{7}{2}b_\m +\ft14\o_\m{}^{ab}\g_{ab})\varphi^i-V_{\m}^{ij}\varphi_j+\tfrac{1}{2} i \slashed{\mathcal{D}}L^{ij}\p_{\m\,j} + \tfrac{1}{2} i \gamma^{a} E_{a} \p_\m^i \nn \\
&&- \tfrac{1}{2} N \p_\m^{i}  + \g \cdot T L^{ij} \p_{\m\,j} - 3 L^{ij}\phi_{\m\,j}\,, \nn \\
\mathcal{D}_\m E_a &=&(\partial_\m-4b_\m)E_a+\o_{\m ab}E^b+\tfrac{1}{2} i \bar{\p}_\m \g_{ab} \mathcal{D}^{b} \varphi  +2 \bar{\p}_\m \gamma^{b} \varphi T_{ba} + 2\bar{\phi}_\m \g_{a} \varphi\,.
\eea
The closure of the superconformal algebra requires that the following constraint must be satisfied
\be
\mathcal{D}^{a} E_{a}= 0\,. \label{closure}
\ee
Thus $E_a$ can be solved in terms of a 3-form $E_{\m\n\r}$ as
\bea
E^{\mu} = - \tfrac{1}{12} \e^{\m\n\r\s\l}\mathcal{D}_\n  E_{\r\s\l},\,
\label{3formE}
\eea
where $E_{\m\n\r}$ is invariant under the following gauge transformation
\bea
\d_\L E_{\m\n\r} = 3 \partial_{[\m} \L_{\n\r ]}\,.
\label{GInvE}
\eea
We can also express $E^{\mu}$ and $E_{\m\n\r}$ in terms of a 2-form potential according to \cite{Coomans:2012cf}
\bea
E^{\m} &=& {\cal{D}}_\n E^{\m\n}\,, \nn\\
E_{\m\n\r} &=& \e_{\m\n\r\s\l} E^{\s\l}\,.
\eea
The supersymmetry transformations of the 2-form gauge field $E_{\m\n}$ and 3-form gauge field $E_{\m\n\r}$ are given in \cite{Coomans:2012cf}
\bea
\d E^{\m\n} &=& - \tfrac12 \rmi \bar\e \g^{\m\n} \vf - \tfrac12 \bar\p_\r^i \g^{\m\n\r} \e^j L_{ij} - \partial_\r \tilde{\L}^{\m\n\r}\,, \nn\\
\d E_{\m\n\r} &=& -  \bar\e \g_{\m\n\r} \vf + \rmi \bar\p_{[\m}^i \g_{\n\r ]} \e^j L_{ij}  \,.
\label{dualityforE}
\eea


\section{Superconformal Actions}\label{s: SCActions}

In this section, we review the derivation of the superconformal action for the linear multiplet \cite{Coomans:2012cf}, and construct a superconformal action for the Yang-Mills multiplet. We begin with the following super-invariant density formula \cite{Fujita:2001kv}
\begin{eqnarray}
e^{-1}{\cal{L}}_{VL} &=& Y^{ij}L_{ij} + i \Bar{\l} \vf - \tfrac{1}{2} \bar{\p}_{a}^{i} \g^{a} \l^{j}L_{ij} -\ft1{12}\epsilon^{\mu\nu\rho\sigma\lambda} A_{\m} \partial_{\nu}E_{\rho\sigma\lambda} \nn\\
&& + \r (N + \tfrac{1}{2} \bar{\p}_{a}\g^{a}\vf + \tfrac{1}{4}i \bar{\p}_{a}^{i} \g^{ab} \p_{b}^{j} L_{ij} ) \,.
\label{VLaction1}
\end{eqnarray}
By integration by parts, ${\cal{L}}_{VL}$ can be reexpressed as
\bea
e^{-1}{\cal{L}}_{VL} &=& Y^{ij}L_{ij} + i \bar{\l} \vf - \tfrac{1}{2} \bar{\p}_{a}^{i} \g^{a} \p^{j}L_{ij} +\tfrac{1}{2} F_{\m\n} E^{\m\n} \nn\\
&& + \r (N + \tfrac{1}{2} \bar{\p}_{a}\g^{a}\vf + \tfrac{1}{4}i \bar{\p}_{a}^{i} \g^{ab} \p_{b}^{j} L_{ij} )\,. \label{VLaction2}
\eea
\subsection{Linear Multiplet Action}
In this section, we construct the vector multiplet in terms of fundamental fields of the linear multiplet and the Dilation Weyl multiplet, and obtain an action for the linear multiplet by using vetor-linear Lagrangian (\ref{VLaction2}). Firstly, the scalar $\r$ in the vector multiplet can be constructed from the elements of linear multiplet as follows \cite{Coomans:2012cf}
\be
\r = 2 L^{-1} N + \rmi \bar{\vf}_i \vf_j L^{ij} L^{-3},\qquad L^2=L^{ij}L_{ij}.
\ee
Using this expression and applying a sequence of supersymmetry transformations, we obtain the full expressions for the components of vector multiplets in terms of elements in the linear multiplet  \cite{Coomans:2012cf}
\bea
\r &=& 2 L^{-1} N + \rmi L^{-3} \bar\vf^i \vf^j L_{ij} \,, \nn\\
\l_i &=& -2 \rmi \slashed{\cal{D}}\vf_i L^{-1} + ( 16 L_{ij}\chi^j + 4 \g \cdot T \vf_i) L^{-1}  - 2 N L_{ij} \vf^j L^{-3}  \nn \\
&& + 2\rmi (\slashed{\cD} L_{ij} L^{jk} \vf_k - \slashed{E} L_{ij} \vf^j ) L^{-3}  + 2 \rmi \vf^j \bar\vf_i \vf_j L^{-3}\nn\\
&& - 6 \rmi \vf^j \bar\vf_k \vf_l L^{kl} L_{ij} L^{-5} ,\nn
\eea
\bea
Y_{ij} &=& L^{-1} \Box^C L_{ij} - {\cal{D}}_{a}L_{k(i} {\cal{D}}^{a}L_{j)m} L^{km} L^{-3} - N^2 L_{ij} L^{-3} - E_{\m}E^{\m}L_{ij} L^{-3}\nn\\
&& + \tfrac{8}{3} L^{-1} T^2 L_{ij} + 4 L^{-1} D L_{ij} + 2 E_{\m} L_{k(i} {\cal{D}}^{\m} L_{j)}{}^{k} L^{-3}  -\ft12 \rmi N L^{-3} \bar\vf_{(i} \vf_{j)} \nn\\
&& - \ft43 L^{-5}N  L_{k(i} L_{j)m} \bar\vf^k \vf^m - \ft23 L^{-3} \bar\vf_{(i} \slashed{E} \vf_{j)} - \ft13 L^{-5} L_{k(i} L_{j)m} \bar\vf^k \slashed{E} \vf^m \nn\\
&& - 8 \rmi   L^{-1} \bar\chi _{(i} \vf_{j)} + 16 \rmi  L^{-3} L_{k(i} L_{j)m} \bar\chi^k \vf^m +2 L^{-3} L_{k(i} \bar\vf^k \slashed{\cD}\vf_{j)} \nn\\
&& + 2 \rmi L^{-3} L_{ij} \bar\vf \g \cdot T \vf - \ft23 L^{-3} \bar\vf_{(i} \slashed{\cD} L_{j)k} \vf^k - L^{-5} L_{mn} L^k{}_{(i} \bar\vf^m \slashed{\cD}L_{j)k}\vf^n \nn\\
&& -\ft16  L^{-5} L_{km} \bar\vf_i \g^a \vf_j \bar\vf^k \g_a \vf^m +\ft1{12} L^{-7} L_{ij} L_{km} L^{pq} \bar\vf^k \g_a \vf^m \bar\vf_p \g^a \vf_q ,\nn\\
\widehat{F}_{\m\n} &=& 4 {\cal{D}}_{[\m}(L^{-1} E_{\n ]}) + 2L^{-1} \widehat{R}_{\m\n}{}^{ij}(V) L_{ij} - 2 L^{-3} L_{k}^{l} {\cal{D}}_{[\m}L^{kp} {\cal{D}}_{\n ]} L_{lp}  \nn\\
&& - 2 \cD_{[\m} (L^{-3} \bar\vf^i \g_{\n]} \vf^j L_{ij} ) - \rmi L^{-1} \bar\vf \widehat{R}_{\m\n} (Q) \,.
\label{embed}
\eea
Substituting above composite expressions into the vector-linear Lagrangian (\ref{VLaction2}), one obtains the superconformal action for the linear multiplet \cite{Coomans:2012cf}
\bea
e^{-1}{\cal{L}}_L &=& L^{-1} L_{ij} \Box^c L^{ij} - L^{ij} {\cal{D}}_{\m}L_{k(i} {\cal{D}}^{\m} L_{j)m} L^{km} L^{-3} + N^2 L^{-1}  \nn\\
&& - E_{\m} E^{\m} L^{-1} + \tfrac{8}{3} L T^{2} + 4 D L - \tfrac{1}{2}L^{-3} E^{\m\n} L_{k}^{l} \partial_{\m} L^{kp} \partial_\n L_{pl}\nn\\
&&  + 2 E^{\m\n} \partial_{\m} ( L^{-1} E_{\n} + V_{\n}^{ij} L_{ij} L^{-1} ) + \text{fermions},
\label{SCAction}
\eea
where the complete expression for the superconformal d'Alembertian is defined as
\bea
L_{ij} \Box^c L^{ij}&=&L_{ij}(\partial^a-4b^a+\omega_b{}^{ba}){\cal{D}}_a L^{ij}+2L_{ij} V_a{}^i{}_k{\cal{D}}^a L^{jk}+6L^2f_a{}^a \nn \\
&&-iL_{ij}\bar{\psi}^{a i}{\cal D}_a\varphi^j-6L^2\bar{\psi}^a\g_a\chi-L_{ij}\bar{\varphi}^i\g\cdot T\g^a\psi_a^j+L_{ij}\bar{\varphi}^i\g^a\phi_a^j\,.
\eea
Fermionic contribution to above action can be straightforwardly read from the formulae given in (\ref{embed}).
\subsection{Yang-Mills Multiplet Action}

 This subsection is devoted to construct a superconformal action describing a vector multiplet coupled to a Dilaton Weyl multiplet. Such an action was previously obtained in \cite{Fujita:2001kv} using the Standard Weyl multiplet and applied in \cite{Hanaki:2006pj} to derive an off-shell Poincar\'e theory. Another use of the vector multiplet action was established in \cite{Bergshoeff:2001hc, Coomans:2012cf}, where the vector multiplet action in the Standard Weyl multiplet background was used to derive the map between the Standard Weyl multiplet and the Dilaton Weyl multiplet.

Similar to the construction of a linear multiplet action, we can use the density formula (\ref{VLaction1}) to obtain an action for an abelian vector multiplet coupled to a Dilaton Weyl multiplet. Based on the action for an abelian vector multiplet, we derive the action for a Yang-Mills multiplet coupled to the Dilaton Weyl Multiplet.

We start from the following identification
\bea
L_{ij} = Y_{ij}.
\eea
This identification, however, has the wrong Weyl weight and fails to satisfy the $S$-invariance of $L_{ij}$ (See Appendix B for the Weyl weights of the fields). The one with the right Weyl weight and invariant under the $S$-transformation is given by
\bea
L_{ij} &=&  \s Y_{ij} + \ft14 \rmi \r \s^{-1} \bar\p_{(i} \p_{j)} - \ft12 \rmi \bar\l_{(i} \p_{j)}.\label{LY}
\eea
After employing a sequence of $Q$-and $S$-transformations to (\ref{LY}), we obtain the full expressions for the components of linear multiplet in terms of the fields in the vector multiplet and Dilaton Weyl multiplet
\bea
L_{ij} &=&  \s Y_{ij} + \ft14 \rmi \r \s^{-1} \bar\p_{(i} \p_{j)} - \ft12 \rmi \bar\l_{(i} \p_{j)}, \nn\\
\vf_i &=& \ft12 \rmi \s \slashed{\cD} \l_i + \ft12 \rmi \r \slashed{\cD} \p_i  +  \r \g \cdot T \p_i +  \s \g \cdot T \l_i  - 8 \s \r \chi_i - \ft18 \g \cdot \widehat G \l_i \nn\\
&&-  \ft18 \g \cdot \widehat F \p_i + \ft14 \slashed{\cD} \s \l_i  + \ft14 \slashed{\cD} \r \p_i - \ft12 Y_{ij} \p^j - \ft1{8} \rmi  \s^{-1} \l^j \bar\p_i  \p_j ,\nn\\
E^{a} &=& \cD_b ( - \ft12 \s \widehat{F}^{ab} - \ft12 \r \widehat{G}^{ab} + 8 \s\r T^{ab} - \ft18 \rmi \bar\l \g_{ab} \p)  - \ft18 \e^{abcde} G_{bc} F_{de}   ,\nn\\
N &=& \ft12 \r \Box^C \s + \ft12 \s \Box^C \r + \ft12 \cD_a \r \cD^a \s - \ft14 \widehat{G}_{ab} \widehat{F}^{ab} - 4 \r \s \Big( {D} + \ft{26}3 T^2\Big)   \nn\\
&& + 4 \s \widehat{F}^{ab} T_{ab} + 4 \r \widehat{G}^{ab} T_{ab} + 8 \rmi \s \bar\chi \l + 8 \rmi \r \bar\chi \p - \ft14  \bar\l \slashed{\cD}\p - \ft14  \bar\p \slashed{\cD}\l \nn\\
&& + \rmi \bar\p \g \cdot T \l.\label{compositeL}
\eea
Inserting above expressions  into density formula (\ref{VLaction1}), we derive an action for an abelian vector multiplet coupled to a Dilaton Weyl multiplet.  Generalization of the action for abelian vector multiplet to that for Yang-Mills multiplet is straightforward. The result is given by
\bea
e^{-1} \cL_{{\rm YM}} &=& a_{IJ} \Big( \s Y_{ij}^I Y^{ij\, J} - \ft14 \s {F}_{\m\n}^I {F}^{\m\n\, J} - \ft12 \r^I  {F}^J_{\m\n} {G}^{\m\n} + 8 \s\r^I {F}^J_{\m\n} T^{\m\n} + \ft12 \r^I \r^J \Box^C \s \nn\\
&& \quad   + \ft12 \s \r^I \Box^C \r^J + \ft12 \r^I \cD_a \r^J \cD^a \s - 4 \s \r^I \r^J (D + \ft{26}3 T^2 ) + 4  \r^I \r^J {G}_{\m\n} T^{\m\n} \nn\\
&& \quad - \ft18  \e^{\m\n\r\s\l} F_{\m\n}^I F_{\r\s}^J C_\l \Big) +\text{fermions},
\label{SCYM}
\eea
where $I = 1, \ldots n$ and the complete expression of the superconformal d'Alembertian for $\r^I$ is \cite{Bergshoeff:2002qk}
\bea
\Box^c \r^I &=& (\partial^a - 2b^a + \o_b{}^{ba}) \cD_a \r^I - \ft12 \rmi \bar\p_a \cD^a \l^I - 2\r^I \bar\p_a \g^a {\chi} \nn\\
&& + \ft12 \bar\p_a \g^a \g \cdot {T} \l^I + \ft12 \bar\phi_a \g^a \l^I + 2 f_a{}^a \r^I.
\eea


\section{Gauge Fixing and Off-Shell Actions}\label{s: GaugeFix}

In the previous section, we have obtained superconformal actions for a linear multiplet and Yang-Mills multiplet. In this section, we fix the redundant superconformal symmetries to obtain off-shell supersymmetric theories including a Poincar\'e supergravity and Yang-Mills coupled to the Dilaton Weyl multiplet.

As discussed in \cite{deWit:1982na}  for the four-dimensional case and in \cite{Hanaki:2006pj} for the five-dimensional case, the construction of a consistent Poincar\'e supergravity requires more than one compensator multiplet if the Standard Weyl multiplet is adopted. However, if the Dilaton Weyl multiplet is utilized, a single compensator multiplet is sufficient to construct a consistent Poincar\'e supergravity. As we will present shortly,  a consistent Poincar\'e supergravity is obtained via gauge fixing the superconformal linear multiplet action \cite{Coomans:2012cf} instead of the vector multiplet action. The latter cannot give rise to a supergravity theory due to the lacking of Einstein-Hilbert term in the action.

In section \ref{ss: GFix} we present our gauge choices and the corresponding decomposition rules. Imposing these gauge choices in the superconformal action, we obtain an off-shell Poincar\'e supergravity in section \ref{ss: Poincare}. In section \ref{ss:Riem}, we first present an off-shell action describing Yang-Mills coupled to a Dilaton Weyl multiplet. Then, using a map between Yang-Mills multiplet and Dilaton Weyl multiplet \cite{Bergshoeff:2011xn}, we obtain an off-shell Riemann tensor squared invariant. Different from \cite{Bergshoeff:2011xn} where the five-dimensional Riemann squared invariant is obtained via circle reduction of the six-dimensional Riemann squared invariant, our construction of the Yang-Mills action is purely based on the superconformal tensor calculus.

\subsection{Gauge Fixing and Decomposition Rules} \label{ss: GFix}
In this section, we introduce our gauge choices to fix the redundant superconformal symmetries in order to obtain off-shell supersymmetric theories. If we do
not insist on the canonical Einstein-Hilbert term in the action, there exists a set of gauge choices facilitating the derivation of curvature squared invariant. These gauge choices are\footnote{The canonical Einstein-Hilbert term can be obtained by using another set of gauge choices \cite{Coomans:2012cf}. }
\be
L_{ij} = \frac{1}{\sqrt{2}}\d_{ij}L, \qquad \s = 1, \qquad \p^i = 0, \qquad b_\m = 0.
\label{gc1}
\ee
The first gauge choice breaks the ${\rm SU}(2)_R$ down to ${\rm U}(1)_R$ whereas the second one fixes dilatations, the third one fixes special supersymmetry transformations and the last one fixes conformal boosts. After fixing the gauge, the remaining fields are
\be
e_\m{}^a(10),\, \p_\m^i(32),\, C_\m(4),\, B_{\m\n}(6),\,\vf^i(8),\,L(1),\, E_{\m\n\r}(4) ,\, N(1),\, V_\m(4) ,\, V_\m^{'ij}(10).
\label{fieldcontent}
\ee

To maintain the gauge (\ref{gc1}), the compensating transformations are required including a compensating ${\rm SU}(2)$, a compensating special supersymmetry and a compensating conformal boost with parameters (up to cubic fermion terms)
\bea
&&\lambda^{ij}=-\frac{1}{\sqrt{2}L}\Big(S^{k(i}\delta^{j)l}\epsilon_{kl}\Big),\quad S^{ij}=\bar{\epsilon}^{(i}\varphi^{j)}-\frac{1}{2}\delta^{ij}\bar{\epsilon}^{k}\varphi^{l}\delta_{kl},\cr
&&\eta^i = \Big( - \g \cdot {T} + \ft14 \g \cdot \widehat G \Big) \e^i,\quad \L_{K\m} = -\ft14 \rmi \bar\e \phi_\m - \ft14 \rmi \bar\eta \p_\m + \bar\e \g_\m {\chi}.
\label{LKm}
\eea

\subsection{Off-Shell Poincar\'e Theory}\label{ss: Poincare}
Imposing the gauge fixing conditions (\ref{gc1}) in the linear multiplet action (\ref{SCAction}), one can obtain a consistent Poincar\'e supergravity whose action is given by\footnote{The action directly coming from (\ref{SCAction}) by imposing (\ref{gc1}) is equal to$-e^{-1}{\cL}_{LR}$.   }
\bea
e^{-1}{\cL}_{LR}	 &=& \tfrac12 L R + \ft12 L^{-1} \partial_\m L \partial^\m L- \ft14 L G_{\m\n} G^{\m\n} - \ft{1}6 L H_{\m\n\r} H^{\m\n\r} - L^{-1} N^2 \nn\\
&&  + \ft16  L^{-1} \partial_{[\m}E_{\n\r\s]} \partial^\m E^{\n\r\s} + \ft1{6\sqrt2} \e^{\m\n\r\s\l} V_\m \partial_\n E_{\r\s\l} +  L V_\m^{'ij} V^{'\m}_{ij}\nn\\
&& +\rm{fermions}.
\label{RLag}
\eea
 Notice that we have decomposed the field $V_\m^{ij}$ into its trace and traceless part as
\bea
V_\m^{ij} = V_\m^{'ij} + \tfrac12 \d^{ij} V_\m, \qquad  V_\m^{'ij} \d_{ij}=0.
\eea
The Poincar\'e supergravity presented above is invariant under the following supersymmetry transformation rules (up to cubic fermion terms)
\bea
\d e_{\m}{}^a &=& \ft12 \bar\e \g^a \p_\m \, ,\nn\\
\d \p_\m^i &=& \cD_\m (\o_{-}) \e^i  - \ft12 \rmi \widehat{G}_{\m\n} \g^\n \e^i \,, \nn\\
\d V_{\m}{}^{ij} &=& \ft12 \bar\e^{(i} \g^\n \widehat\p_{\m\n}^{j)} - \ft16 \bar\e^{(i} \g \cdot \widehat{H} \p_\m^{j)} - \ft14 \rmi \bar\e^{(i} \g \cdot \widehat{G} \p_\m^{j)}+\partial_{\m}\lambda^{ij}+\lambda^{(i}_{~k}V^{j)k}_{\mu},  \nn\\
\d C_\m &=& -\ft12\rmi\bar\e \p_\m \, ,\nn\\
\d B_{\m\n} &=& \ft12 \bar\e \g_{[\m} \p_{\n]} + C_{[\m} \d(\e) C_{\n]} \,,\nn\\
\d L &=& \ft1{\sqrt2} \rmi \bar\e^i \vf^j \d_{ij} \,,\nn\\
\d \vf^i &=& - \ft1{2\sqrt2} \rmi \slashed{\partial} L \d^{ij} \e_j - \ft1{\sqrt2} \rmi V'_{\m}{}^{(i}{}_k \d^{j)k} L \e_j  - \ft12 \rmi \slashed{E} \e^i  + \ft12 N \e^i + \ft1{4\sqrt2} L \g \cdot \widehat{G} \d^{ij} \e_j \, \nn\\
&&  - \ft1{6\sqrt2} \rmi L \g \cdot \widehat{H} \d^{ij} \e_j  ,\nn\\
\d E_{\m\n\r} &=& - \bar\e \g_{\m\n\r} \vf + \ft1{\sqrt2} \rmi L \bar\p^i_{[\m} \g_{\n\r ]} \e^j \d_{ij} \,, \nn\\
\d N &=& \ft12 \bar\e \g^\m \Big( \partial_\m + \ft14 \o_\m{}^{bc} \g_{bc} \Big) \vf + \ft12 \bar\e^i \g^a V_{a\,ij} \vf^j - \ft1{4\sqrt2} \rmi \bar\e^i \g^a \slashed{\partial}L  \p_a^j \d_{ij}\nn\\
&& + \ft1{4\sqrt2}\rmi \bar\e^i \g^a \g^b V'_{b(i}{}^k \d_{j)k} \p_a^j + \ft14 \rmi \bar\e \g^a \slashed{E} \p_{a} - \ft14 N \bar\e  \g^a \p_a + \ft1{8\sqrt2} L \bar\e^i \g^a \g \cdot \widehat{G} \p_a^j \d_{ij}\nn\\
&&-\sqrt2 L \bar\e^i \g^a \phi_a^j \d_{ij} + \ft18 \rmi \bar\e \g \cdot \widehat{H} \vf ,
\label{UnGaugedTransform}
\eea
where we have used the torsionful spin connection \cite{Bergshoeff:2011xn}
\bea
{\omega}_\mu{}^{ab}_\pm &=& {\omega}_\mu{}^{ab} \pm {\widehat
H}_\mu{}^{ab}\ ,
\label{torsion} \eea
and the supercovariant curvatures under the gauge (\ref{gc1}) are \cite{Bergshoeff:2011xn}
\bea
{\widehat\psi}_{\mu\nu} &=& 2D_{[\mu} ({\omega}_-) \psi_{\nu]} +i\g^\lambda
\widehat G_{\lambda[\mu}\psi_{\nu]}\ ,
\label{psiab}\\
\widehat{G}_{\m\n} &=& 2\partial _{[\mu }C_{\nu]} + \ft12\rmi \bar{\p}_{[\m} \p_{\n]}\ ,
\label{ghat}\\
\widehat{H}_{\m\n\r} &=& 3\partial _{[\mu }B_{\nu \rho ]} - \ft34
\bar{\p}_{[\m} \g_\n \p_{\r]} + \ft32 C_{[\m} G_{\n\r]}\ .
\label{hhat}
\eea

\subsection{Riemann Squared Action}\label{ss:Riem}

In this section, we construct the supersymmetric Riemann squared action. To begin with, we shall review a map between the Yang-Mills super-multiplet and a set of fields in the Poincar\'e multiplet (\ref{fieldcontent}).

In establishing the map between Yang-Mills and Poincar\'e multiplets, it is important to consider the full supersymmetry transformations, including the cubic fermion terms which have been omitted so far. In the following, we shall need the full supersymmetry transformation rules for the fields $(e_\m{}^a, \p_\m^i,V^{ij}_{\mu}, C_\m,B_{\m\n})$. Up to cubic fermions, the transformation rules of $(e_\m{}^a, \p_\m^i,V^{ij}_{\mu}, C_\m,B_{\m\n})$ are already given in (\ref{UnGaugedTransform}). In this section, we will, however, keep the complete ${\rm SU}(2)$ symmetry, i.e. we do not impose $L^{ij}=\ft1{\sqrt{2}}L\delta^{ij}$. In this way we do not need to accommodate for the compensating  ${\rm SU}(2)$ transformations proportional to $\lambda^{ij}$\footnote{After we construct the action, we can still impose the gauge $L^{ij}=\ft1{\sqrt{2}}L\delta^{ij}$. This will not affect the Riemann squared invariant.}. The full version of the supersymmetry transformations are given by \cite{Bergshoeff:2011xn}
\begin{eqnarray}\label{fixed}
\d e_\m{}^a  & = &   \ft 12\bar\e \g^a \psi_\m \ , \nn\\
\d \psi_\m^i  & = & D_\m (\omega_-) \e^i -\ft12 \rmi
{\widehat G}_{\mu\nu} \g^\nu\e^i \ , \nn\\
\d V_\m{}^{ij} & = & \ft12 \bar\e^{(i} \g^\nu \psi_{\mu\nu}^{j)}
-\ft16 \bar\e^{(i} \g\cdot {\widehat H}\psi_\mu^{j)} -\ft14 \rmi
\bar\e^{(i} \g\cdot {\widehat G}\psi_\mu^{j)}\ , \nn\\
\d C_\m & = & -\ft12\rmi \bar{\e} \p_\m \ , \nn\\
\d B_{\m\n} & = & \ft12  \bar{\e} \g_{[\m} \p_{\n]} + C_{[\m} \d(\e)
C_{\n]}\,.
\end{eqnarray}
Next, we consider the following supersymmetry transformations \cite{Bergshoeff:2011xn}
\begin{eqnarray}
\delta \o_\mu{}^{ab}_+ &=& -\frac12\rmi\widehat
G^{ab}\bar\e\psi_\mu -\ft12\bar\e\g_\mu\widehat{\psi}^{ab}, \nn\\
\d {\widehat\psi}^i_{ab} &=& \tfrac14 \g^{cd} {\widehat R}_{cdab}
({\omega}_+) \e^i - {\widehat V}_{ab}{}^{ij}\e_j + \ft12 i
\gamma^\mu {\cD}_\mu (\o_+) \widehat{G}_{ab} \e^i -
\tfrac14 \widehat{G}_{ab} \g \cdot \widehat{G} \e^i, \nn\\
\delta \widehat G_{ab} &=&  -\frac12 \rmi
\bar\e \widehat{\psi}_{ab},\nn\\
\d {\widehat V}_{ab}{}^{ij} &=& -\tfrac12 \bar{\e}^{(i}
\slashed{\cD}({\o}, {\o}_-)
{\widehat\psi}^{j)}_{ab} - \tfrac{1}{24} \bar{\e}^{(i} \g \cdot
\widehat{H} \psi^{j)}_{ab} - \rmi \bar{\e}^{(i} \widehat{G}^d{}_{[a}
{\widehat\psi}_{b]d}^{j)}\, ,\label{curtrans}
\end{eqnarray}
where ${\widehat R}_{abcd} ({\omega}_+)$ denotes the
super-covariant curvature of the torsionful connection
${\o}_+$. In ${\cD}_\mu (\o_+)
\widehat{G}_{ab}$, the connection ${\o}_+$ rotates both the indices $a$ and $b$, and  in
${\cD}_\mu({\o},{\o}_-){\widehat\psi}^{j}_{ab}$
the connection ${\o}$ rotates the spinor index, while the
connection ${\o}_-$ rotates the Lorentz vector indices. ${\widehat V}_{\mu\nu}{}^{ij}$ is the supercovariant curvature of $V^{ij}_{\m}$ under the gauge choices (\ref{gc1})
\bea {\widehat V}_{\mu\nu}{}^{ij} &=& V_{\mu\nu}{}^{ij} -
{\bar\psi}_{[\mu}^{(i}\gamma^\rho {\widehat\psi}_{\nu]\rho}^{j)} +\ft1{6} {\bar\psi}_\mu^{(i} \gamma\cdot {\widehat
H}\psi_\nu^{j)} +\ft14 i {\bar\psi}_\mu^{(i} \gamma\cdot {\widehat
G} \psi_\nu^{j)}\ . \label{vmn}
\eea
We now compare the above transformation rules with those of the $D=5$, $ \cN =2 $ Yang-Mills multiplet \cite{Bergshoeff:2011xn}
\begin{eqnarray}
\d A_\m^I &=& -\ft12\rmi \rho^I \bar{\e} \p_\m + \ft12 \bar{\e}
\g_\m \lambda^I \ ,
\nonumber \\
\d Y^{ij\, I} &=& -\ft12 \bar{\e}^{(i} {\slashed{\cD}} \lambda^{j) I} -
\tfrac{1}{24} \bar{\e}^{(i} \g \cdot \widehat{H} \lambda^{j)I} -
\ft12 i g \bar{\e}^{(i} f_{JK}{}^I \rho^J \lambda^{j)K} \ ,
\nonumber \\
\d \lambda^{i I} &=& - \ft14 \left(\g \cdot \widehat{F}^I-\rho^I \g\cdot \widehat{G}\right)
\e^i -\ft12\rmi {\slashed{\cD}}\rho^I \e^i - Y^{ij\, I} \e_j \ ,
\nonumber \\ \d \rho^I &=& \ft12
\rmi \bar{\e} \lambda^I \ , \label{vectPoincare}
\end{eqnarray}
where $\widehat{F}_{\m\n}^I$ and ${\cD}_\mu\, \rho^I$ can be found in \eqref{hatF} and \eqref{deriv1} by imposing the gauge choices (\ref{gc1})
\bea
{\cD}_\mu \lambda^{i I} &=&
(\partial_\mu + \ft14\,  \o_\mu
{}^{ab}\g_{ab} ) \l^{iI} - V_\mu^{ij} \l_j^I + g f_{JK}{}^I A_\mu^J
\lambda^{iK}
\nn\\
&& + \ft14 \left( \g
\cdot {\widehat{F}}^I - \rho^I \g\cdot {\widehat G}\right) \p_\mu^i
+ \ft12\rmi {\slashed{\cD}}\rho^I  \p_\mu^i  + Y^{ij\, I} \p_{\mu\, j}\ .
\eea
We observe that the transformations (\ref{curtrans}) and (\ref{vectPoincare}) become identical by making the following identifications \cite{Bergshoeff:2011xn}
\bea
( A_\m^I, \quad Y_I^{ij}, \quad \l_I^i, \quad \r_I) \quad \longleftrightarrow \quad ( \o_{\m +}^{ab}, \quad - \widehat{V}_{ab}{}^{ij}, \quad - \widehat\p_{ab}^i, \quad \widehat{G}_{ab} ).
\label{YMDWMap}
\eea
Setting $a_{IJ} = \d_{IJ}$ and imposing the gauge fixing conditions (\ref{gc1}) in action (\ref{SCYM}), we obtain
\bea
e^{-1} \cL_{{\rm YM} }|_{\s = 1}&=& Y_{ij}^I Y^{ij I} - \ft12 D_\m \r^I D^\m \r^I  -\ft14 (F_{ab}^I - \r^I G_{ab}) (F^{ab I} - \r^I G^{ab}) \nn\\
&& - \ft18 \e^{abcde} (F^I_{ab} - \r^I G_{ab})(F_{cd}^I - \r^I G_{cd}) C_e \nn\\
&& - \ft12 \e^{abcde} (F_{ab}^I - \r^I G_{ab}) B_{cd} D_e \r^I+{\rm fermions}.
\label{preRie1}
\eea
Finally, using the map (\ref{YMDWMap}) in above action (\ref{preRie1}), we obtain the supersymmetric Riemann squared action. Its purely bosonic part is given as
\bea  e^{-1}{\cal L}_{{\rm Riem}^2}  &=& -\ft14
\Big(\,R_{\mu\nu ab}(\omega_+)- G_{\mu\nu} G_{ab}\Big)
\left(\,R^{\mu\nu ab}(\omega_+)- G^{\mu\nu} G^{ab}\right) \nn\\ &&
-\ft12 \nabla_\mu(\omega_+) G^{ab} \nabla^\mu(\omega_+)G_{ab}
+V_{\mu\nu}{}^{ij} V^{\mu\nu}{}_{ij} \nn\\ && - \ft1{8}
\e^{\mu\nu\rho\sigma\lambda} \Big(\,R_{\mu\nu
ab}(\omega_+)- G_{\mu\nu} G_{ab}\Big)
\left(\,R_{\rho\sigma}{}^{ab}(\omega_+)- G_{\rho\sigma}
G^{ab}\right) C_\lambda \nn\\ && - \ft12
\e^{\mu\nu\rho\sigma\lambda} B_{\rho\sigma}\left(\,R_{\mu\nu
ab}(\omega_+)- G_{\mu\nu}G_{ab}\right) \nabla_\lambda (\omega_+)
G^{ab}+{\rm fermions}\,.
\label{bosonicR2}
 \eea
We notice that the actions (\ref{preRie1}) and (\ref{bosonicR2}) obtained via superconformal tensor calculus match with those derived through the circle reduction of six-dimensional actions \cite{Bergshoeff:2011xn}.

\section{Supersymmetric Gauss-Bonnet Combination} \label{s: SCGB}

In this section, we shall construct the supersymmetric completion of the Gauss-Bonnet combination
\bea
e^{-1} \cL_{{\rm GB}} &=& R_{\m\n\r\s} R^{\m\n\r\s} - 4 R_{\m\n} R^{\m\n} + R^2.
\eea
According to the usual routine, one may think of constructing three independent curvature squared super-invariants first, then combining them with proper coefficients to form a supersymmetric Gauss-Bonnet combination. However, as we mentioned before, two independent curvature squared invariants may be enough to obtain the supersymmetric completion of Gauss-Bonnet combination based on counting the degrees of freedom and the cancelation of the kinetic term for the auxiliary vector $V^{ij}_{\mu}$. This section is devoted to construct another curvature squared invariant.

We start from the conventional constraint imposed on the supercovariant curvature of $\o_\m{}^{ab}$ \cite{Fujita:2001kv,Bergshoeff:2001hc}
\be
e^\n{}_b \widehat{R}_{\m\n}{}^{ab}(M) = 0,\label{constrant1}
\ee
where $\widehat{R}_{\m\n}{}^{ab}(M)$ is defined in (\ref{RM}). The conventional constraint (\ref{constrant1}) implies that the supercovariant curvature of $\o_{\m}{}^{ab}$ gives the Weyl Tensor, which is defined as
\bea
C_{\m\n\r\s} &=& R_{\m\n\r\s} - \ft13 (g_{\m\r}R_{\n\s} - g_{\n\r} R_{\m\s} - g_{\m\s} R_{\n\r} + g_{\n\s} R_{\m\r} ) \nn\\
&& + \ft1{12} (g_{\m\r} g_{\n\s} - g_{\m\s} g_{\n\r} ) R.
\eea
Its square is
\bea
C_{\m\n\r\s} C^{\m\n\r\s} &=& R_{\m\n\r\s} R^{\m\n\r\s} - \ft43 R_{\m\n} R^{\m\n} + \ft16 R^2.
\eea
In the rest of this paper, we use $\widehat{C}_{\mu\nu\rho\sigma}$ to denote the superconformally covariant Weyl tensor instead of $\widehat{R}_{\m\n}{}^{ab}(M)$.
Because the off-shell supersymmetric Riemann squared invariant is known, the Gauss-Bonnet super-invariant can be obtained by combining the Riemann squared invariant with another curvature squared invariant in which the curvature squared terms take the form
\be
C_{\m\n\r\s} C^{\m\n\r\s} + \ft16 R^2\label{target}.
\ee
Although, none of the terms in
 (\ref{target}) is a supercovariant quantity, we can replace (\ref{target}) by the following supercovariant expression
\be
\widehat{C}_{\m\n\r\s}  \widehat{C}^{\m\n\r\s} + \ft{512}3 D^2,
\label{Key}
\ee
since the composite field $D$ (\ref{UMap}) under the gauge choices (\ref{gc1}) reads
\be
D=-\ft1{32}R-\ft1{16}G^{ab}G_{ab}-\ft{26}3T^{ab}T_{ab}+2T^{ab}G_{ab}+{\rm fermions}\label{DD}.
\ee
Therefore, if (\ref{Key}) can be supersymmetrized, we will get the desired the curvature squared terms in (\ref{target}). When carrying out the supersymmetrization of
(\ref{Key}), we find that in fact, the $D^2$ term is indispensable to the supersymmetrization of the Weyl tensor squared term,
moreover, the relative coefficient between the Weyl squared term and the  $D^2$ exactly matches with the one in (\ref{Key}), the magical $\ft{512}{3}$. In the next section, we give the details of the construction.

\subsection{ Supersymmetrization of $\widehat{C}_{\m\n\r\s}  \widehat{C}^{\m\n\r\s}$}
In this section, we first supersymmetrize the square of Weyl tensor by using (\ref{VLaction2}) in which the fields of linear multiplet are expressed as
composites in terms of fields in Dilaton Weyl multiplet. We notice that to obtain the Weyl tensor squared term, $N$ should begin with $\widehat{C}_{\m\n\r\s}  \widehat{C}^{\m\n\r\s}$. The complete expression for $N$ include a term $\ft{512}{3}D^2$. After expanding $D$ in terms of independent fields, we find that the curvature squared terms take the form of $C^{\mu\nu\rho\sigma} C_{\mu\nu\rho\sigma} + \ft16 R^2$, which is different from those in the supersymmetric completion of $\widehat{C}_{\m\n\r\s}  \widehat{C}^{\m\n\r\s}$ considered in \cite{Hanaki:2006pj} by using Standard Weyl multiplet where $D$ is merely an auxiliary field. We obtain full composite expressions for the fields of linear multiplet in terms of fields in the Dilation Weyl multiplet as
\bea
L^{ij} &=&  \ft14 \rmi \bar{\hat{R}}^{(i}_{ab}(Q) \hat{R}^{j) ab} (Q) + \ft{256}3 \rmi {\bar\chi}^{(i} {\chi}^{j)} + \ft{16}3 \widehat{R}_{ab}{}^ {ij}(V) {T}^{ab},\nn\\
\vf^i &=& - \ft1{8} \g_{cd} \widehat{R}_{ab}^i(Q) \widehat{C}^{abcd} - 4 \rmi \g_c \widehat{R}_{ab}^i(Q) \cD^a {T}^{bc}+ \ft{128}3 {\chi}^i D  \nn\\
&&  + 8 \rmi \g_c \cD^c \widehat{R}_{ab}^i(Q) {T}^{ab} + 8 \rmi \g_a \cD^c \widehat{R}_{bc}^i(Q) {T}^{ab}- \ft{64}3 \rmi \g^{ab} \g^c \cD^a {T_{bc}}\, {\chi}^i + \ft{1024}9 {T}^2\, {\chi}^i \nn\\
&& + \ft{128}3 \rmi \g_a \cD_b {\chi}^i {T}^{ab} + \ft{16}3 \g_{ab} \widehat{R}_{cd}^i(Q) {T}^{ab} {T}^{cd} + \ft12 \widehat{R}^{ab\, i}{}_j(V) \widehat{R}_{ab}^j(Q) \nn\\
&& - \ft83 \widehat{R}^{ab\, i}{}_j(V) \g_{ab} {\chi}^j, \nn\\
E_a &=& \ft1{16} \e_{abcde} C^{bcfg} C^{de}{}_{fg} -\ft1{12} \e_{abcde} V^{bc}{}_{ij} V^{de\, ij}  \nn\\
&& + \cD^b \Big(4C_{abcd} {T}^{cd} - \ft{64}3 {D} \, {T_{ab}} - \ft{128}9 {T_{ab}} {T}^2 - \ft{512}3 {T_{ac}} {T}^{cd} {T_{bd}} \Big) \nn\\
&&  -32 \e_{abcde} \cD^b \Big( \ft{2}3  {T}^{cf} \cD_f {T}^{de} +{T}^c{}_f \cD^d {T}^{ef} \Big)+{\rm fermions}  , \nn\\
N &=& \ft18 C^{abcd} C_{abcd} + \ft{64}3 {D}^2 + \ft{1024}9 {T}^2 {D} - \ft{16}3C_{abcd} {T}^{ab} \, {T}^{cd} - \ft{1}3 V_{ab}{}^{ij}V^{ab}{}_{ij} \nn\\
&& - \ft{64}3 \cD_a T_{bc} \cD^a T^{bc} + \ft{64}3 \cD_b T_{ac} \cD^a T^{bc}- \ft{128}3 {T_{ab}} \cD^b \cD_c {T}^{ac}\nn\\
&& - \ft{128}3 \e_{abcde} {T}^{ab} {T}^{cd} \cD_f {T}^{ef} + 1024 \, {T}^4 - \ft{2816}{27}({T}^2)^2+{\rm fermions}.\label{embedLtoD}
\eea
where the following notations are introduced for simplicity
\bea
T^4 \equiv T_{ab} T^{bc} T_{cd} T^{da}, \qquad (T^2)^2 \equiv (T_{ab} T^{ab})^2.
\eea
Under the gauge choices (\ref{gc1}) $T_{ab} \cD^b \cD_c T^{ac}$ is given by
\be
T_{ab} \cD^b \cD_c T^{ac} = T_{ab} \nabla^b \nabla_c T^{ac} + \ft23 R^{bc} T_{ab} T^a{}_c - \ft1{12} T^2 R+{\rm fermions},
\ee
where $\nabla_{\mu}$ only contains the usual spin connection
\bea
\nabla_\m T_{ab} &=& \partial_\m T_{ab} - 2 \o_\m{}^c{}_{[a} T_{b]c}.
\eea
To obtain (\ref{embedLtoD}) we have used the $Q$- and $S$- transformations of supercovariant curvatures which can be found in \cite{Bergshoeff:2001hc}.
Substituting the composite expressions  (\ref{embedLtoD}) into the density formula (\ref{VLaction2}), we obtain the following action
\bea
e^{-1} \cL_{\r R^2} &=&  \ft18\r {C}^{abcd} {C}_{abcd}+ \ft{64}3 \r {D}^2 + \ft{1024}9 \r {T}^2 {D} - \ft{32}3 {D} \, {T_{ab}} F^{ab}   \nn\\
&&  - \ft{16}3 \r {C}_{abcd} {T}^{ab} \, {T}^{cd} + 2{C}_{abcd} {T}^{cd} F^{ab} + \ft1{16} \e_{abcde}A^a {C}^{bcfg} {C}^{de}{}_{fg}    \nn\\
&& -\ft1{12} \e_{abcde} A^a {V}^{bc}{}_{ij} {V}^{de\, ij} +  \ft{16}3 Y_{ij} {V}_{ab}{}^ {ij} {T}^{ab} - \ft{1}3 \r {V}_{ab}{}^{ij}{V}^{ab}{}_{ij}+\ft{64}3 \r  \cD_b T_{ac} \cD^a T^{bc} \nn\\
&& - \ft{128}3 \r {T_{ab}} \cD^b \cD_c {T}^{ac} - \ft{64}3  \r \cD_a T_{bc} \cD^a T^{bc} + 1024 \r \, {T}^4- \ft{2816}{27} \r  ({T}^2)^2   \nn\\
&&- \ft{64}9 {T_{ab}} F^{ab} {T}^2 - \ft{256}3 {T_{ac}} {T}^{cd} {T_{bd}} F^{ab}  - \ft{32}3   \e_{abcde}  {T}^{cf} \cD_f {T}^{de} F^{ab}  \nn\\
&& - 16   \e_{abcde} {T}^c{}_f \cD^d {T}^{ef} F^{ab}  - \ft{128}3 \r \e_{abcde} {T}^{ab} {T}^{cd} \cD_f {T}^{ef} + \text{fermions},
\label{pregb}
\eea
where
\be
V_{\m\n}{}^{ij} \equiv 2\partial_{[\mu}V_{\nu]}{}^{ij} -2V_{[\mu}{}^{k(i}V_{\nu]k}{}^{j)}.
\ee
This action (\ref{pregb}) describes the coupling between an external vector multiplet and Dilaton Weyl multiplet. If we simply combine above action with the Riemann tensor squared invariant, we are not able to obtain the supersymmetric Gauss-Bonnet combination since the curvature squared terms in (\ref{pregb}) is multiplied by $\r$ which stays the same after imposing the gauge choices (\ref{gc1}).
By comparing the superconformal transformation rules of vector multiplet
\bea
\d\r &=& \ft12 \rmi \bar\e \l ,\nn\\
\d A_\m &=& - \ft12 \s \bar\e \p_\m + \ft12 \bar\e \g_\m \l ,\nn\\
\d \l^i &=& -\ft14 \g \cdot \widehat F \e^i - \ft12 \rmi \slashed{\cD}\r \e^i + \r \g \cdot {T} \e^i  - Y^{ij} \e_j + \r \eta^i ,\nn\\
\d Y^{ij} &=& - \ft12 \bar\e^{(i} \slashed{\cD} \l^{j)} + \ft12 \rmi \bar\e^{(i} \g \cdot {T} \l^{j)}  - 4 \rmi \r \bar\e^{(i} {\chi}^{j)} + \ft12 \rmi \bar\eta^{(i} \l^{j)},
\eea
with those of $(\s,\,C_\m,\,\p^i)$ in the Dilaton Weyl multiplet
\bea
\d\s &=& \ft12 \rmi \bar\e \p,\nn\\
\d C_\m &=& - \ft12 \s \bar\e \p_\m + \ft12 \bar\e \g_\m \p, \nn\\
\d \p^i &=& -\ft14 \g \cdot \widehat G \e^i - \ft12 \rmi \slashed{\cD}\s \e^i + \s \g \cdot {T} \e^i - \ft14 \rmi \s^{-1} \e_j \bar\p^i \p^j + \s \eta^i,
\eea
we notice that there exists a map from vector multiplet to $(\s,\,C_\m,\,\p^i)$
\be
\r \rightarrow \s, \qquad A_a \rightarrow C_a, \qquad \l^i \rightarrow \p^i,  \qquad  Y^{ij} \rightarrow \ft14 \rmi \s^{-1} \bar\p^{(i} \p^{j)},
\label{vtodw}
\ee
 since
 \be
\d (\ft14 \rmi \s^{-1} \bar\p^{(i} \p^{j)}) = - \ft12 \bar\e^{(i} \slashed{\cD} \p^{j)} + \ft12 \rmi \bar\e^{(i} \g \cdot {T} \p^{j)}  - 4 \rmi \s \bar\e^{(i} {\chi}^{j)} + \ft12 \rmi \bar\eta^{(i} \p^{j)}.
\ee

Using (\ref{vtodw}), we obtain the supersymmetrization of $C_{\m\n\r\s} C^{\m\n\r\s}$ purely based on the fields of Dilaton Weyl multiplet
\bea
e^{-1} \cL_{\sigma C^2} &=&  \ft18\s {C}^{abcd} {C}_{abcd}+ \ft{64}3 \s {D}^2 + \ft{1024}9 \s {T}^2 {D} - \ft{32}3 {D} \, {T_{ab}} G^{ab}   \nn\\
&&  - \ft{16}3 \s {C}_{abcd} {T}^{ab} \, {T}^{cd} + 2{C}_{abcd} {T}^{cd} G^{ab} + \ft1{16} \e_{abcde}C^a {C}^{bcfg} {C}^{de}{}_{fg}    \nn\\
&& -\ft1{12} \e_{abcde} C^a {V}^{bc}{}_{ij} {V}^{de\, ij} +  \ft{16}3 Y_{ij} {V}_{ab}{}^ {ij} {T}^{ab} - \ft{1}3 \s {V}_{ab}{}^{ij}{V}^{ab}{}_{ij}+\ft{64}3 \s  \cD_b T_{ac} \cD^a T^{bc} \nn\\
&& - \ft{128}3 \s {T_{ab}} \cD^b \cD_c {T}^{ac} - \ft{64}3  \s \cD_a T_{bc} \cD^a T^{bc} + 1024 \s \, {T}^4- \ft{2816}{27} \s  ({T}^2)^2   \nn\\
&&- \ft{64}9 {T_{ab}} G^{ab} {T}^2 - \ft{256}3 {T_{ac}} {T}^{cd} {T_{bd}} G^{ab}  - \ft{32}3   \e_{abcde}  {T}^{cf} \cD_f {T}^{de} G^{ab}  \nn\\
&& - 16   \e_{abcde} {T}^c{}_f \cD^d {T}^{ef} G^{ab}  - \ft{128}3 \s \e_{abcde} {T}^{ab} {T}^{cd} \cD_f {T}^{ef} + \text{fermions}.
\label{prepregb2}
\eea
Imposing the gauge fixing conditions (\ref{gc1}), we obtain
\bea
e^{-1} \cL_{\sigma C^2}|_{\s =1} &=& \ft18  R_{abcd} R^{abcd} - \ft16  R_{ab} R^{ab} + \ft1{48}  R^2 + \ft{64}3 D^2 + \ft{1024}9  {T}^2 {D}\nn\\
&&  - \ft{16}3  {R}_{abcd} {T}^{ab} \, {T}^{cd} + 2{R}_{abcd} {T}^{cd} G^{ab} + \ft13 R T_{ab} G^{ab} - \ft83 R_{bd} G_c{}^b T^{cd} \nn\\
&&  - \ft{64}3 R^{bc} T_{ab} T^a{}_c  +\ft{8}3 R T^2  - \ft{32}3 {D} \, {T_{ab}} G^{ab}  + \ft1{16} \e_{abcde}C^a {R}^{bcfg} {R}^{de}{}_{fg} \nn\\
&& -\ft1{12} \e_{abcde} C^a V^{bc}{}_{ij} V^{de\, ij}  - \ft{1}3 V_{ab}{}^{ij} V^{ab}{}_{ij} -\ft{64}3   \nabla_a T_{bc} \nabla^a T^{bc} \nn\\
&&+ \ft{64}3  \nabla_b T_{ac} \nabla^a T^{bc} - \ft{128}3  {T_{ab}} \nabla^b \nabla_c {T}^{ac} - \ft{128}3 \e_{abcde} {T}^{ab} {T}^{cd} \nabla_f {T}^{ef} \nn\\
&&+ 1024  \, {T}^4- \ft{2816}{27}  ({T}^2)^2  - \ft{64}9 {T_{ab}} G^{ab} {T}^2 - \ft{256}3 {T_{ac}} {T}^{cd} {T_{bd}} G^{ab}   \\
&&  - \ft{32}3   \e_{abcde}  {T}^{cf} \nabla_f {T}^{de} G^{ab}- 16   \e_{abcde} {T}^c{}_f \nabla^d {T}^{ef} G^{ab}+{\rm fermions},\nn
\label{pregb2}
\eea
where
\bea
D&\equiv&-\ft1{32}R-\ft1{16}G^{ab}G_{ab}-\ft{26}3T^{ab}T_{ab}+2T^{ab}G_{ab}+{\rm fermions},\cr
T_{ab} &\equiv& \ft18 G_{ab} + \ft1{48} \e_{abcde} H^{cde}+{\rm fermions}.
\eea
\subsection{Supersymmetric Completion of Gauss-Bonnet Combination}
In previous sections, we obtained the supersymmetric completion of Einstein-Hilbert, Riemann tensor squared and Weyl tensor squared actions. Because of the off-shell nature of these invariants, we can combine them to form
a more general theory with two free parameters
\be
{\cal L}={\cal L}_{LR}+\alpha{\cal L}_{{\rm Riem}^2}+\beta\cL_{\sigma C^2}\mid_{\s =1}\label{gaction}.
\ee
 The Gauss-Bonnet
combination corresponds to case with $\beta=3\alpha$ in which the kinetic term of auxiliary vector $V^{ij}_{\mu}$ vanishes.
Using $\beta=3\alpha$, the purely bosonic part of Lagrangian (\ref{gaction}) takes the form
\bea
e^{-1} \Big({\cal L}_{LR}+\alpha{\cal L}_{{\rm GB}}\Big) &=& \tfrac12 L R + \ft12 L^{-1} \partial_\m L \partial^\m L- \ft14 L G_{\m\n} G^{\m\n} - \ft{1}6 L H_{\m\n\r} H^{\m\n\r} - L^{-1} N^2 \nn\\
&&   + \ft16  L^{-1} \partial_{[\m}E_{\n\r\s]} \partial^\m E^{\n\r\s} + \ft1{6\sqrt2} \e^{\m\n\r\s\l} V_\m \partial_\n E_{\r\s\l} +  L V_\m^{'ij} V^{'\m}_{ij}\nn\\
&& + \a \Big[ -\ft14
\Big(\,R_{\mu\nu ab}(\omega_+)- G_{\mu\nu} G_{ab}\Big)
\Big(\,R^{\mu\nu ab}(\omega_+)- G^{\mu\nu} G^{ab}\Big) \nn\\
&& + \ft38   R_{\m\n\r\s} R^{\m\n\r\s} - \ft12   R_{\m\n} R^{\m\n} + \ft1{16}   R^2 + {64}  D^2\nn\\
&& - \ft1{8}
\e^{\mu\nu\rho\sigma\lambda} \Big(\,R_{\mu\nu
ab}(\omega_+)- G_{\mu\nu} G_{ab}\Big)
\Big(\,R_{\rho\sigma}{}^{ab}(\omega_+)- G_{\rho\sigma}
G^{ab}\Big) C_\lambda \nn\\
 && - \ft12
\e^{\mu\nu\rho\sigma\lambda} B_{\rho\sigma}\Big(\,R_{\mu\nu
ab}(\omega_+)- G_{\mu\nu}G_{ab}\Big) \nabla_\lambda (\omega_+)
G^{ab} \nn\\
&&  + \ft3{16} \e_{\m\n\r\s\l}C^\m {R}^{\n\r\t\d} {R}^{\s\l}{}_{\t\d} -\ft1{4} \e_{\m\n\r\s\l} C^\m V^{\n\r}{}_{ij} V^{\s\l\, ij} \nn\\
&&  - {16} {R}_{\m\n\r\s} {T}^{\m\n} \, {T}^{\r\s} + 6 {R}_{\m\n\r\s} {G}^{\m\n} T^{\r\s} +  R T_{\m\n} G^{\m\n} - 8 R_{\m \n} G_\s{}^\m T^{\s\n} \nn\\
&&  - 64 R^{\m\n} T_{\s\m} T^\s{}_\n  + 8 R T^2  - 32 {D} \, {T_{\m\n}} G^{\m\n}  + \ft{1024}3  {T}^2 {D}   - 64   \nabla_\m T_{\n\r} \nabla^\m T^{\n\r} \nn\\
&&+ {64}   \nabla^\m T^{\n\r}  \nabla_\n T_{\m\r} - {128} {T_{\m\n}} \nabla^\n \nabla_\s {T}^{\m\s}  -\ft12  \nabla_\mu(\omega_+) G^{ab} \nabla^\mu(\omega_+)G_{ab}  \nn\\
&&+ 3072  \, {T}^4- \ft{2816}{9}  ({T}^2)^2  - \ft{64}3 {T_{\m\n}} G^{\m\n} {T}^2 - {256} {T_{\m\s}} {T}^{\s\r} {T_{\r\n}} G^{\n\m}   \nn\\
&&  - {128} \e_{\m\n\r\s\l} {T}^{\m\n} {T}^{\r\s} \nabla_\t {T}^{\l\t}  - {32}   \e_{\m\n\r\s\l}  G^{\m\n} {T}^{\r\t} \nabla_\t {T}^{\s\l} \nn\\
&&-  48   \e_{\m\n\r\s\l} G^{\m\n} {T}^\r{}_\t \nabla^\s {T}^{\l\t}  \Big].
\label{GaussB}
\eea
We notice that the ratio of the coefficients in front of the Gauss-Bonnet combination and the Chern-Simons coupling $\e_{\mu\nu\rho\sigma\lambda}C^\m {R}^{\n\r\delta\tau} {R}^{\sigma\lambda}{}_{\delta\tau}$ is $\frac{1}{2}$ which is consistent with the value resulting from the circle reduction of the partial results given in \cite{Bergshoeff:1986vy,Bergshoeff:1986wc} on the six-dimensional supersymmetric Gauss-Bonnet combination.

\section{On-Shell Theory}

In this section, we study the on-shell theory of the Gauss-Bonnet extended supergravity
to first order in $\a$ upon eliminating the auxiliary fields. In section \ref{ss:OnP}, we present the minimal on-shell Poincar\'e supergravity by eliminating the the auxiliary fields $(E_{\m\n\r}, V_\m, N, V_\m{}^{'ij})$ and truncating the matter multiplet $(B_{\m\n}, L, \vf^i)$. In section \ref{ss:OnGB}, we obtain the on-shell Gauss-Bonnet extended Einstein-Maxwell supergravity to first order in $\a$ by using the equations derived from the 2-derivative Lagrangian that is zeroth order in $\a$.

\subsection{On-Shell Poincar\'e Supergravity}\label{ss:OnP}

To eliminate the auxiliary fields $(N, P_a, V_\m, V_\m{}^{'ij})$, we use their equations of motion
\bea
\label{eqnaux1}
0 &=& N,\quad 0= \e^{\m\n\r\s\l} \partial_\n E_{\r\s\l}, \qquad  0 = V_\m^{'ij}\,, \\
0 &=&  \partial^\m ( L^{-1} \partial_{[\m} E_{\n\r\s]} + \frac{1}{2\sqrt2} \e_{\m\n\r\s\l} V^\l ) \,.
\label{eqnaux2}
\eea
Equation (\ref{eqnaux2}) implies that locally
\bea
- \frac{1}{12} L^{-1} \e^{\m\n\r\s\l} \partial_\n E_{\r\s\l} + \frac{1}{\sqrt2} V_\m &=& \partial_\m \phi \,,
\eea
where $\phi$ is a Stueckelberg scalar. Eliminating this scalar by using the shift symmetry transformation and using the second equation in (\ref{eqnaux1}), we obtain \footnote{In the original Poincar\'e theory (\ref{RLag}) ${\rm U}(1)_R$ symmetry is gauged by the auxiliary vector $V_\m$. However, in the on-shell theory, the ${\rm U}(1)_R$ symmetry becomes global due to the elimination of $V_{\mu}$.}
\bea
V_\m = 0 \,.
\eea
 It follows that the corresponding on-shell theory is given by
\bea
e^{-1}{\cL'}_{{\rm EM}} &=& \frac12 L R + \frac12 L^{-1} \partial_\m L \partial^\m L- \frac14 L G_{\m\n} G^{\m\n} - \frac{1}6 L H_{\m\n\r} H^{\m\n\r} \,.
\label{DualEM}
\eea
To truncate out the matter multiplet $(B_{\m\n}, L, \vf^i)$, we first dualize $B_{\m\n}$ to a vector field $\tilde{C}_{\m}$ by adding the following Lagrange multiplier to (\ref{DualEM})
\be
\D \cL = -\frac{1}{12} \e^{\m\n\r\s\l} B_{\m\n\r} \widetilde{G}_{\s\l},\quad \widetilde{G}_{\m\n} \equiv 2 \partial_{[\m} \tilde{C}_{\n]},
\ee
and replacing $H_{\mu\nu\rho}$ by $B_{\m\n\r}+\ft32C_{[\m}G_{\nu\rho]}$. The field equations of $\tilde{C}_{\m}$ and $B_{\mu\nu\rho}$ imply that
\be
B_{\m\n\r}=3\partial_{[\m}B_{\nu\rho]},\quad H^{\m\n\r} = - \frac14 L^{-1} \e^{\m\n\r\s\l} \widetilde{G}_{\s\l} \,.
\label{dualH}
\ee
Substituting (\ref{dualH}) to (\ref{DualEM}), we obtain the on-shell ungauged Einstein-Maxwell supergravity
\bea
e^{-1}{\cL}_{{\rm EM}} &=& \frac12 L R + \frac12 L^{-1} \partial_\m L \partial^\m L- \frac14 L G_{\m\n} G^{\m\n} - \frac18 L^{-1} \widetilde{G}_{\m\n}  \widetilde{G}^{\m\n} \nn\\
&& + \frac18  \e^{\m\n\r\s\l} C_\m G_{\n\r}  \widetilde{G}_{\s\l} \,,
\eea
where $(e_\m{}^a, \p_\m^i, C_\m )$ constitute the supergravity multiplet while $(\tilde{C}_\m, \vf^i, L)$ comprise the Maxwell multiplet.

Truncation of the Einstein-Maxwell theory to the minimal on-shell theory can be implemented by imposing
\be
L = 1,\quad \widetilde{C}_\m = C_\m,\quad \vf^i = 0,\label{truncation}
\ee
which is consistent with the equations of motion
\bea
R &=& 2 L^{-1} \Box L - L^{-2} \partial_\m L \partial^\m L + \ft12 G_{\m\n} G^{\m\n} - \ft14 L^{-2} \widetilde{G}_{\m\n} \widetilde{G}^{\m\n}\,, \nn\\
L R_{\m\n} &=& \nabla_\m \nabla_\n L - L^{-1}\partial_\m L \partial_\n L + L G_\m{}^\s G_{\n\s} + \ft12 L^{-1} \widetilde{G}_\m{}^\s \widetilde{G}_{\n\s} \nn\\
&& - \ft14 g_{\m\n} L^{-1} \widetilde{G}_{\r\s} \widetilde{G}^{\r\s} \,,\cr
0&=&\nabla^{\nu}(L G_{\n\m})+\frac14\epsilon_{\mu\nu\rho\sigma\lambda}G^{\nu\rho}\widetilde{G}^{\sigma\lambda},\cr
0&=&\nabla^{\nu}(L^{-1}\widetilde{ G}_{\n\m})+\frac14\epsilon_{\mu\nu\rho\sigma\lambda}G^{\nu\rho}G^{\sigma\lambda},
\eea
and leads to the following transformation
\bea
\d e_\m{}^a &=& \ft12 \bar\e \g^a \p_\m \,, \nn\\
\d \p_\m^i &=& ( \partial_\m + \ft14 \o_\m{}^{ab} \g_{ab} ) \e^i + \ft18 \rmi ( \g_\m{}^{\n\r} - 4 \d_\m^\n \g^\r ) G_{\n\r},\, \nn\\
\d C_\m &=& - \ft12 \rmi \bar\e \p_\m.
\eea
The resulting action coincides with the minimal on-shell supergravity in five dimensions \cite{Gunaydin:1983bi}
\bea
e^{-1} \cL_{{\rm EH}}^{{\rm min}} =  \frac12 R - \frac38 G_{\m\n} G^{\m\n}  + \frac18  \e^{\m\n\r\s\l} C_\m G_{\n\r} G_{\s\l} \,.
\eea
The canonical kinetic term of $C_{\m}$ can be recovered by a scaling $C_\m\rightarrow\frac{2}{\sqrt{ 6}} C_\m$.

\subsection{On-Shell Gauss-Bonnet Extended Einstein-Maxwell Model} \label{ss:OnGB}

With the Gauss-Bonnet combination added, the duality relation (\ref{dualH}) and truncation condition must receive corrections proportional to the powers of $\alpha$, if we consider a pertubative expansion valid when the energy scale $\Lambda$ satisfies $\Lambda^2\ll 1/|\alpha|$. We follow the procedure of \cite{Argyres:2003tg}. Schematically, the off-shell action (\ref{GaussB}) takes the form
\bea
S_{\text{off-shell}} [\phi] &=& S_0 [\phi] + \a S_1 [\phi] \,.
\label{expS}
\eea
It follows that the auxiliary field equations (\ref{eqnaux1}) - (\ref{eqnaux2}), the field equation for $B_{\m\n\r}$ (\ref{dualH}) as well as the truncation equation (\ref{truncation}) must receive corrections proportional to $\a$. The solution to those equations can be expressed in terms of a series expansion in $\a$
\bea
\phi &=& \phi_0 + \a \phi_1 + \a^2 \phi_2 + \cdots \,,
\eea
where $\phi_0$  is the solution to the zeroth order equation given in previous section. As a consequence, the on-shell action possesses the form
\bea
S_{\text{on-shell}} [\phi] &=& S_0 [\phi_0] + \a (S_1 [\phi_0]+\phi_1 S'_0[\phi_0])+ \cdots\,.
\eea
In the above equation, $S'_0[\phi_0]=0$ when $\phi_0$ is an auxiliary field or a Lagrangian multiplier. We eliminate the auxiliary fields and Lagrangian multiplier $B_{\m\n\r}$ by plugging their zeroth order solutions to the action (\ref{GaussB}). Ultimately we derive the on-shell Gauss-Bonnet extended Einstein-Maxwell theory\footnote{Generalization of the Gibbons-Hawking boundary term in theories with generic curvature-squared corrections in the presence of a chemical potential is studied in \cite{Cremonini:2009ih}.}
\bea
&&e^{-1} \Big({\cal L}_{{\rm EM}}+\alpha{\cal L}_{{\rm GB}}\Big)= \frac12 L R + \frac12 L^{-1} \partial_\m L \partial^\m L- \frac14 L G_{\m\n} G^{\m\n} - \frac18 L^{-1} \widetilde{G}_{\m\n}  \widetilde{G}^{\m\n}    \nn\\
&&\qquad\qquad + \frac18  \e^{\m\n\r\s\l} C_\m G_{\n\r}  \widetilde{G}_{\s\l}+ \a \Big[\ft38   R_{\m\n\r\s} R^{\m\n\r\s} - \ft12   R_{\m\n} R^{\m\n} + \ft1{16}   R^2  \nn\\
&&\qquad\qquad+{64}  D^2  -\ft14
\Big(\,R_{\mu\nu ab}(\omega_+)- G_{\mu\nu} G_{ab}\Big)
\Big(\,R^{\mu\nu ab}(\omega_+)- G^{\mu\nu} G^{ab}\Big)\nn\\
&&\qquad\qquad - \ft1{8}
\e^{\mu\nu\rho\sigma\lambda} \Big(\,R_{\mu\nu
ab}(\omega_+)- G_{\mu\nu} G_{ab}\Big)
\Big(\,R_{\rho\sigma}{}^{ab}(\omega_+)- G_{\rho\sigma}
G^{ab}\Big) C_\lambda \nn\\
 &&\qquad\qquad - \ft12
\e^{\mu\nu\rho\sigma\lambda} B_{\rho\sigma}\Big(\,R_{\mu\nu
ab}(\omega_+)- G_{\mu\nu}G_{ab}\Big) \nabla_\lambda (\omega_+)
G^{ab}  + \ft3{16} \e_{\m\n\r\s\l}C^\m {R}^{\n\r\t\d} {R}^{\s\l}{}_{\t\d}  \nn\\
&&\qquad\qquad - {16} {R}_{\m\n\r\s} {T}^{\m\n} \, {T}^{\r\s} + 6 {R}_{\m\n\r\s} {G}^{\m\n} T^{\r\s} +  R T_{\m\n} G^{\m\n} - 8 R_{\m \n} G_\s{}^\m T^{\s\n} \nn\\
&&\qquad \qquad- 64 R^{\m\n} T_{\s\m} T^\s{}_\n  + 8 R T^2  - 32 {D} \, {T_{\m\n}} G^{\m\n}  + \ft{1024}3  {T}^2 {D}   - 64   \nabla_\m T_{\n\r} \nabla^\m T^{\n\r} \nn\\
&&\qquad\qquad+ {64}   \nabla^\m T^{\n\r}  \nabla_\n T_{\m\r} - {128} {T_{\m\n}} \nabla^\n \nabla_\s {T}^{\m\s}  -\ft12  \nabla_\mu(\omega_+) G^{ab} \nabla^\mu(\omega_+)G_{ab}  \nn\\
&&\qquad\qquad+ 3072  \, {T}^4- \ft{2816}{9}  ({T}^2)^2  - \ft{64}3 {T_{\m\n}} G^{\m\n} {T}^2 - {256} {T_{\m\s}} {T}^{\s\r} {T_{\r\n}} G^{\n\m}   \nn\\
&&\qquad\qquad  - {128} \e_{\m\n\r\s\l} {T}^{\m\n} {T}^{\r\s} \nabla_\t {T}^{\l\t}  - {32}   \e_{\m\n\r\s\l}  G^{\m\n} {T}^{\r\t} \nabla_\t {T}^{\s\l} \nn\\
&&\qquad\qquad-  48   \e_{\m\n\r\s\l} G^{\m\n} {T}^\r{}_\t \nabla^\s {T}^{\l\t}  \Big]+{\cal O}(\alpha^2),
\label{EMGB}
\eea
where $T_{\m\n}$ and $\o_{+\m}{}^{ab}$ are now given by
\be
T_{\m\n} = \ft1{16} ( 2 G_{\m\n} + L^{-1} \widetilde{G}_{\m\n} ),\quad\o_{+\m}{}^{ab} =\o_\m{}^{ab} - \ft14 L^{-1} e_{\m f} \e^{fabcd} \widetilde{G}_{cd}.
\ee

\section{Vacuum Solutions and Spectrum Analysis} \label{s: Vacuum}
In this section, we investigate the vacuum solutions and spectrum to the general theory (\ref{gaction}). The results for Poincar\'e supergravity extended by Gauss-Bonnet combination can be obtained as special case when $\beta=3\alpha$.

\subsection{Vacuum Solutions with 2-form and 3-form Fluxes}
We first consider solutions with $AdS_3\times S^2$ structure. To solve the equation of motion, we make the following ansatz where Greek indices denote the coordinates on Lorentzian $AdS_3$, while latin indices stand for the coordinates on $S^2$
\bea
&&R_{\mu\nu\rho\sigma}=-a(g_{\mu\rho}g_{\nu\sigma}-g_{\mu\sigma}g_{\nu\rho}),\quad
R_{pqrs}=b(g_{pr}g_{qs}-g_{ps}g_{qr}),\cr
&&L=L_0,\qquad G_{pq}=c\epsilon_{pq},\qquad H_{\mu\nu\rho}=d\epsilon_{\mu\nu\rho}.\label{ansatz1}
\eea
In above equation, $\varepsilon_{\mu\nu\rho}$ and $\varepsilon_{rs}$ are the Levi-Civita tensors
on $AdS_3$ and $S^2$ respectively.
The full set of equations of motion are solved provided that the following equations are satisfied
\bea
&&6 a-2 b+c^2-2 d^2=0,\cr
&&\ft12L_0(-a+d^2)+\frac{\alpha}{2}(-a^2 + b^2 - 2 b c^2 + c^4 - 4 a c d + 10 a d^2 + 4 c d^3 - 9 d^4)\cr
&& \qquad\qquad\qquad~+\frac{\beta}{6}(a^2 + a b - b^2 + 2 b c^2 - c^4 + 2 a c d - 10 a d^2 - b d^2 -
   2 c d^3 + 9 d^4)=0,\cr
&&\ft14L_0(b - c^2)+\frac{\alpha}{2}(3 a^2 - b^2 + 4 b c^2 - 3 c^4 - 4 b c d + 4 c^3 d - 6 a d^2 +
   3 d^4)\cr
&&\qquad\qquad\qquad~+\frac{\beta}{6} (-3 a^2 + b^2 - 4 b c^2 + 3 c^4 + 4 b c d - 4 c^3 d + 6 a d^2 -
   3 d^4)=0.
\eea
The integrability conditions for the Killing spinor equations $\delta_{\epsilon}\psi^i_{\mu}=0$ and
$\delta_{\epsilon}\varphi^i=0$ are
\be
\Big(R_{\hat{\mu}\hat{\nu}\hat{a}\hat{b}}(\omega_-)-2G_{\hat{\mu}\hat{a}}G_{\hat{\nu}\hat{a}}\Big)\gamma^{\hat{a}\hat{b}}\epsilon=0,
\qquad \Big(\ft32G_{\hat{\mu}\hat{\nu}}-i H_{\hat{\mu}\hat{\nu}\hat{\lambda}}\gamma^{\hat{\lambda}}\Big)\gamma^{\hat{\mu}\hat{\nu}}\epsilon=0,\label{Integra}
\ee
where $\hat{\mu},\,\hat{a}=0,1,\ldots 4$.
Substituting the ansatz (\ref{ansatz1}) into the integrability conditions (\ref{Integra}), we find
that when
\be
a=d^2,\qquad b=c^2,\qquad c=-2d,
\ee
the integrability conditions are satisfied automatically without imposing any projection condition on the
$Q$ transformation parameter $\epsilon$. Therefore, this solution possesses maximum supersymmetry. Remarkably, this solution exists for arbitrary values of $L_0,\,\alpha,\,\beta$. Thus it seems that
the higher derivative correction will not affect the supersymmetric solutions. A similar phenomenon happens
in 6D chiral gauged supergravity extended by Riemann squared invariant \cite{Bergshoeff:2012ax}.
Next we investigate solutions with $AdS_2\times S^3$ structure. We make similar ansatz as previous case
except that Greek indices denote the coordinates on Lorentzian $AdS_2$, while latin indices are used for the coordinates on $S^3$
 \bea
&&R_{\mu\nu\rho\sigma}=-b(g_{\mu\rho}g_{\nu\sigma}-g_{\mu\sigma}g_{\nu\rho}),\quad
R_{pqrs}=a(g_{pr}g_{qs}-g_{ps}g_{qr}),\cr
&&L=L_0,\qquad G_{\mu\nu}=c\epsilon_{\mu\nu},\qquad H_{pqr}=d\epsilon_{pqr}.
\eea
In this case, the solutions of equation of motion are determined by
\bea
&&6 a-2 b+c^2-2 d^2=0,\cr
&&\ft12L_0(a-d^2)+\frac{\alpha}{2}(-a^2 + b^2 - 2 b c^2 + c^4 - 4 a c d + 10 a d^2 + 4 c d^3 - 9 d^4)\cr
&& \qquad\qquad\qquad~+\frac{\beta}{6}(a^2 + a b - b^2 + 2 b c^2 - c^4 - 2 a c d - 10 a d^2 - b d^2 +
   2 c d^3 + 9 d^4)=0,\cr
&&\ft14L_0(-b + c^2)+\frac{\alpha}{2}(3 a^2 - b^2 + 4 b c^2 - 3 c^4 - 4 b c d + 4 c^3 d - 6 a d^2 +
   3 d^4)\cr
&&\qquad\qquad\qquad~+\frac{\beta}{6} (-3 a^2 + b^2 - 4 b c^2 + 3 c^4 - 4 b c d + 4 c^3 d + 6 a d^2 -
   3 d^4)=0.
\eea
By examining the integrability conditions (\ref{Integra}), we find that solution with maximum supersymmetry is given by
\be
a=d^2,\qquad b=c^2,\qquad c=2d,
\ee
for arbitrary values of $L_0,\,\alpha,\,\beta$.
\subsection{Vacuum Solutions Without Fluxes}
If we set $c=d=0$, the solutions are simply
\bea
&&1)\quad AdS_3\times S^2: b=3a,\quad \beta=6\alpha,\quad a=-\frac{L_0}{2\alpha},\cr
&&2)\quad AdS_2\times S^3: b=3a,\quad \beta=6\alpha,\quad a=\frac{L_0}{2\alpha},\cr
&&3)\quad \rm{Minkowski}_5
\eea
In this case, the maximally supersymmetric vacuum solution is just $\rm{Minkowski}_5$. Following the procedure carried out in the spectrum analysis of six-dimensional higher derivative chiral supergravity \cite{Bergshoeff:2012ax,Pang:2012xs}, we study the
bosonic spectrum of the perturbations around the maximally supersymmetric $\rm{Minkowski}_5$ vacuum.
We define the linearized fluctuations,
\bea
&&g_{\mu\nu}=\eta_{\mu\nu}+h_{\mu\nu},\qquad L=L_0+\phi,\qquad C_{\mu}=c_{\mu},\cr
&&V_{\mu}^{ij}=v^{ij}_{\mu},\qquad B_{\mu\nu}=b_{\mu\nu}.
\eea
The linearized Einstein equation and $L$ field equation
take the following form
\bea
\Big(L_0+\ft23(\beta-3\alpha)\Box\Big)R^{(L)}_{\mu\nu}&=&\ft13(\beta-3\alpha)\partial_{\mu}\partial_{\nu}R^{(L)}
+\ft{L_0}{2}\eta_{\mu\nu}R^{(L)}-\eta_{\mu\nu}\Box\phi+\partial_{\mu}\partial_{\nu}\phi,\label{greom}\\
L_0 R^{(L)}&=&2\Box\phi,\label{sceom}
\eea
where $R^{(L)}_{\mu\nu}$ and $R^{(L)}$ are the linearized Ricci tensor and Ricci scalar.
Inserting (\ref{sceom}) into the trace of linearized Einstein equation, we get
\be
\Big( L_0+\ft23(\beta-3\alpha)\Box\Big)\Box\phi=0\label{ss}.
\ee
This equation describes a massless scalar and a massive scalar with mass squared
\be
m^2=\frac{3L_0}{2(3\alpha-\beta)}.
\ee
To simplify the linearized Einstein equation, we choose the usual De Donder gauge in which,
\be
R^{(L)}_{\mu\nu}=-\ft12\Box h_{\mu\nu}.
\ee
Then using the (\ref{sceom}) and (\ref{ss}), we find
\be
(\Box-m^2)\Box h_{\mu\nu}=-2L_0^{-1}(\Box-m^2)\partial_{\mu}\partial_{\nu}\phi.
\ee
Since $\phi$ can be solved from (\ref{ss}), the right hand side of above equation is known function. The homogeneous solutions of above equation describe a massless graviton and a massive graviton with a mass squared the same as that of the massive scalar.

Equations of motion for the remaining fields can be straightforwardly obtained by choosing the
Lorentz gauge for the gauge fields
\be
\Big( L_0+\ft23(\beta-3\alpha)\Box\Big)\Box\left(
                                         \begin{array}{c}
                                           c_{\mu} \\
                                           b_{\mu\nu}\\
                                         \end{array}
                                       \right)
=0,\quad
( L_0+\ft23(\beta-3\alpha)\Box\Big)v_{\mu}^{ij}=0.
\ee
In summary, for generic $\alpha,\,\beta$, the full spectrum consists of the (reducible) massless 12+12 supergravity multiplet with fields
$(h_{\mu\nu},\,b_{\mu\nu},\,c_{\mu},\,\phi,\,\psi^{i}_{\mu},\,\varphi^i)$ and a massive 32+32 supergravity
multiplet with ghost fields $(h_{\mu\nu},\,b_{\mu\nu},\,c_{\mu},\,\phi,\,v^{ij}_{\mu},\psi^{i}_{\mu},\,\varphi^i)$.
At the special point where $\beta=3\alpha$, the curvature squared terms in the action furnish the Gauss-Bonnet combination,
massive particles become infinitely heavy and decouple from the spectrum leaving only the massless excitations as expected from the ghost-free feature of Gauss-Bonnet combination.

\section{Conclusion and Discussions}\label{s:Conc}

Using the superconformal tensor calculus in five dimensions, we have constructed an off-shell
theory with four parameters \bea
e^{-1} \cL &=& {\cal L}_{LR} + \xi\cL_{ {\rm YM}}|_{\s=1} +\alpha{\cal L}_{{\rm Riem}^2}+\beta{\cal L}_{\s C^2}|_{\s=1}  + \zeta \cL_{\r R^2} |_{\s =1}.
\label{total}
\eea
The supersymmetric Gauss-Bonnet extended Poincar\'e theory corresponds to the case where $\xi = \zeta = 0$ and  $ \b = 3 \a$. Although the auxiliary fields do not propagate in this model, they can be eliminated order by order in $\alpha$. We obtain the on-shell theory of this model to first order in $\a$. The maximally supersymmetric solutions to the ordinary 2-derivative Einstein-Maxwell supergravity are known including  ${\rm Minkowski}_5, AdS_3 \times S^2$ and $AdS_3 \times S^2$. We found that these solutions are not modified by the inclusion of the higher-derivative interactions proportional to $\alpha$ and $\beta$ for arbitrary values. The spectrum of this theory around the maximally supersymmetric
${\rm Minkowski}_5$ is determined. We show that the spectrum has a ghostly massive spin two multiplet in addition to a massless supergravity and a Maxwell vector multiplet. However, when $\beta=3\alpha$ corresponding to the Gauss-Bonnet combination, the massive spin-2 multiplet decouples.

Our off-shell model is ungauged and therefore does not admit $AdS_5$ as a supersymmetric vacuum solution. The gauging of our model should be interesting.
A further question is the matter couplings of this theory. Since neither ``very special geometry" \cite{Gunaydin:1983bi, deWit:1992cr}, nor ``quaternionic K\"ahler geometry" \cite{Bergshoeff:2002qk} arise naturally in our model via the gauge fixing condition (\ref{gc1}), it would be interesting to investigate how the scalars in the vector multiplet and hypermultipet are constrained and what kind of geometries arise.
Finally, we hope to generalize our construction to $D=6,\, \cN = (1,0)$ off-shell supergravity to derive the supersymmetric completion of the Gauss-Bonnet combination in six dimensions.

\section*{Acknowledgements}
We thank Frederik Coomans, Chris Pope for useful discussions and Ergin Sezgin for his careful reading and useful remarks which help us improve this paper. Y.P. is supported in part by DOE grant DE-FG03-95ER40917.

\appendix
\section{Notations and Conventions}

In this paper, we use the conventions of \cite{Bergshoeff:2001hc}. The signature of the metric is diag$(-, +, +, +, +)$. The SU(2) indices are lowered or raised according to NW-SE convention
\be
A^i = \ve^{ij} A_j, \qquad A_i = A^j \ve_{ji},
\ee
where $\ve_{12} = - \ve_{21} = \ve^{12} = 1$. When SU(2) indices on spinors are suppressed, NW-SE contraction is understood.
\be
\bar\p \g^{a_1 \ldots\, a_n} \chi = \bar\p^i\g^{a_1 \ldots\, a_n} \chi_i,
\ee
where $\g^{a_1 \ldots a_n}$ is defined as
\be
\g^{a_1 \cdot a_n} = \g^{[a_1} \g^{a_2 \, \ldots} \g^{a_n ]}.
\ee
Changing the order of spinors in a bilinear leads to the following signs
\be
\bar\p^i \g_{(n)} \chi^j = t_n \bar\chi^j \g_{(n)} \p^i,
\ee
where $t_0 = t_1 = - t_2 = - t_3 = 1$. We also used the following Fierz identity
\be
\p_j \bar\chi^i = - \ft14 \bar\chi^i \p_j - \ft14 \bar\chi^i \g^a \p_j \g_a + \ft18 \bar\chi^i \g^{ab} \p_j \g_{ab}.
\ee
The Levi-Civit\'a tensor is real and satisfies
\be
\e_{p_1\ldots p_n q_1\ldots q_m}\e^{p_1\ldots p_n r_1\ldots r_m} = - n! m! \d_{[q_1 \ldots}^{[r_1 \ldots} \d_{q_m]}^{r_m]}.
\ee
Finally, the product of all gamma matrices is proportional to the unit matrix, and we use
\be
\g^{abcde} = \rmi \e^{abcde}.
\ee

\section{Multiplets of Five Dimensional Superconformal Gravity}
In this appdendix, we give the ${\rm SU}(2)$ representations and Weyl weights of the fields appearing in this paper.
\begin{center}
\begin{tabular}{ c | c c c }
Multiplet & Field & ${\rm SU}(2)$ reps. & Weyl weight\\
\hline \hline
Dilaton Weyl Multiplet & $e_\m{}^a$ & 1 & -1 \\
                                   & $\p_\m^i$  & 2 & $-\ft12$ \\
                                   & $b_\m$  & 1 & $ 0 $ \\
                                   & $V_\m^{ij}$  & 3 & $ 0$ \\
                                   & $C_\m$  & 1& $ 0 $ \\
                                   & $B_{\m\n}$  & 1 & $ 0 $ \\
                                   & $\s$  & 1 & $ 1 $ \\
                                   & $\p^i$  & 2 & $\ft32$ \\
\hline \hline
Vector Multiplet & $A_\m $ & 1 &  0 \\
                                   & $\l^i$  & 2 & $\ft32$ \\
                                   & $\r$  & 1 & $ 1 $ \\
                                   & $Y^{ij}$  & 3 & $ 2$ \\
\hline \hline
Linear Multiplet & $L^{ij} $ & 3 &  3 \\
                                   & $\vf^i$  & 2 & $\ft72$ \\
                                   & $E_a$  & 1 & $4  $ \\
                                   & $N$  & 1& $ 4$
\end{tabular}
\end{center}

\end{document}